\documentclass[10pt]{article}
\usepackage[a4paper,top=2.54cm,bottom=2.54cm,left=3.18cm,right=3.18cm]{geometry}
\usepackage{graphicx,color,xcolor,subcaption}
\usepackage{amsmath,amssymb,amsthm,mathtools,amsfonts,mathrsfs,bm}
\usepackage{algorithm,listings}
\usepackage{algpseudocode}
\usepackage{titlesec}
\usepackage{xr-hyper}
\usepackage{hyperref,url}
\usepackage{authblk}
\usepackage[version=4]{mhchem}

\usepackage{float}

\usepackage{lipsum}
\newtheorem*{theorem}{THEOREM}

\title{Stochastic Kinetics of mRNA Molecules\\ in a General Transcription Model}

\author{Yuntao Lu \thanks{Email: \href{yuntaolu22@m.fudan.edu.cn}{yuntaolu22@m.fudan.edu.cn}}}

\author{Yunxin Zhang \thanks{Corresponding author. Email: \href{xyz@fudan.edu.cn}{xyz@fudan.edu.cn}}}

\affil{School of Mathematical Sciences, Fudan University,\\ Shanghai 200433, CHINA}

\date{\today}

\begin{document}

\maketitle
\begin{abstract}
Stochastic modeling of transcription is a classic yet long-standing problem in theoretical biophysics. The lack of unified results and a computationally efficient approach for a general, fine-grained transcription model has confined relevant research to some over-simplified special cases like the Telegraph model. This article establishes a general, unified and computationally efficient framework for studying stochastic transcription kinetics. We consider a chemical reaction model of transcription and construct the time-dependent solution to the corresponding chemical master equation. A well-known matrix-form expression for steady-state binomial moments is recovered by calculating the temporal limit of the time-dependent dynamics. Two novel inequalities for binomial moments and the probability mass function are derived using techniques from functional analysis. It follows that the distribution of mRNA counts is upper-bounded by a constant multiple of Poisson distribution, thus mathematically proving the main statement of the Heavy-Tailed Law. Additionally, the standard binomial moment method is analyzed from a numerical perspective, where truncation error is estimated using our inequalities. Compared with some widely-used numerical methods, a key advantage of this result is the significantly lower computational complexity.
\newline

\textbf{Significance}: This article develops a general, unified and computationally efficient framework for analyzing stochastic transcription kinetics. While pursuing the utmost theoretical generality of the model, we derive the exact time-dependent solution to the corresponding chemical master equation, but only the steady-state distribution is numerically tractable using the standard binomial moment method. The result unifies many existing studies and features lower computational complexity. We also derive two novel inequalities for binomial moments and the probability mass function, and demonstrate their utility in several applications.
\end{abstract}

\tableofcontents

\newpage

\section{Introduction}
Stochastic modeling of gene expression is a classic yet long-standing problem in theoretical biophysics. Transcription, the synthesis of messenger RNA (mRNA) directed by genetic information encoded in DNA, typically constitutes the first step of gene expression and is fundamental to almost all cellular processes. 
Notably, intrinsic stochasticity is a key feature of transcription. The stochasticity stems from random collisions among molecules, and becomes dominant due to low copy numbers of mRNA and the typically single-copy nature of DNA loci in individual cells \cite{stochasticvsdeterministic2024Nat.Commun.,Centraldogmasinglemolecule2011Nature}. This intrinsic fluctuation contributes prominently to the phenotypic heterogeneity \cite{NoiseGeneExpression2005Science}, and has been studied extensively both experimentally \cite{MechanismTranscriptionalBursting2014Cell} and mathematically \cite{StochasticProcessesCell2021,Summingnoisegene2004Nature,Modelsstochasticgene2005Phys.LifeRev.}.

Given the stochasticity and discreteness of mRNA molecules, the temporal evolution of mRNA copy number during transcription should be modeled as a continuous-time stochastic process with a discrete state space. 
The simplest class of such stochastic processes is the continuous-time Markov chain, whose Kolmogorov forward equation is also termed the chemical master equation (master equation or CME) \cite{AppliedStochasticAnalysis2019,StochasticProcessesPhysics2007} in chemical physics literature. A classical mathematical description of transcription is the Telegraph model, where the gene switches memorylessly between one active and one inactive state \cite{MarkovianModelingGeneProduct1995Theor.Popul.Biol.}. The Telegraph model is widely used to quantitatively study the bursting kinetics \cite{UsingGeneExpression2012Science,MammalianGenesAre2011Science} and transcriptional regulation \cite{DistanceMattersImpact2013Phys.Rev.Lett.}, mainly because it admits a concise and computationally tractable expression for the steady-state distribution of mRNA counts. However, the Telegraph model is an oversimplification of transcription \cite{Whatcanwe2024PLoSComput.Biol.}.
Substantial experimental evidence suggests that transcription is a complex process governed by intricate mechanism \cite{MechanismTranscriptionalBursting2014Cell}. In particular, holding time of inactive state is observed to be nonexponential \cite{MammalianGenesAre2011Science}, implying the complicated structure of gene states. Thereby, numerous refined mathematical models of transcription have been proposed and theoretically analyzed \cite{Exactdistributionsstochastic2022Phys.Rev.E,Exactlysolvablemodels2020J.Chem.Phys.,StochasticGeneExpression2019SIAMJ.Appl.Math.,MultimodalityFlexibilityStochastic2013Bull.Math.Biol.,AnalyticalTimeDependentDistributions2023SIAMJ.Appl.Math.,StochasticityTranscriptionalRegulation2001Biophys.J.,NascentRNAkinetics2024Phys.Rev.E,PromotermediatedTranscriptionalDynamics2014Biophys.J.}.

Essentially, the challenge stems from the intractability of CMEs. 
Closed-form, time-dependent solutions exist only for a few trivial cases. Consequently, CMEs are typically studied approximately \cite{Approximationinferencemethods2017J.Phys.A-Math.Theor.} or numerically \cite{StochasticSimulationChemical2007Annu.Rev.Phys.Chem.,finitestateprojection2006J.Chem.Phys.}. Among approximation methods, the diffusion approximation \cite{AppliedStochasticAnalysis2019}, which approximates the CME with the chemical Langevin equation, is the most widely used. However, the chemical Langevin equation generates continuous sample paths, which is often inappropriate in stochastic transcription models because some cells contain extremely few mRNA molecules and discreteness cannot be ignored. Among numerical methods, the stochastic simulation algorithm (also known as Gillespie algorithm or SSA) \cite{generalmethodnumerically1976J.Comput.Phys.} and the finite state projection algorithm (FSP) \cite{finitestateprojection2006J.Chem.Phys.} are two widely-used, general-purpose approaches. SSA is a Monte Carlo algorithm that generates sample paths of a given continuous-time Markov chain one at a time. Numerous refinements of the original SSA have been developed \cite{StochasticSimulationChemical2007Annu.Rev.Phys.Chem.}. Nevertheless, despite their broad applicability, stochastic simulation methods are computationally expensive in general, and quickly become impractical for large-scale reaction systems. FSP approximates the solution to a CME by direct truncation, reducing the problem to solving an ordinary differential equation system or evaluating a matrix exponential. Theoretical results are available to estimate truncation errors \cite{finitestateprojection2006J.Chem.Phys.}, but FSP still becomes infeasible for large systems because computing the matrix exponential is expensive. Additionally, approaches based on queueing theory are also exploited to provide theoretical results for a broad class of stochastic transcription models \cite{IntrinsicNoiseStochastic2011Phys.Rev.Lett.,Inferringtranscriptionalbursting2023R.Soc.OpenSci.,QueuingModelsGene2020Biophys.J.,Solvingstochasticgeneexpression2024Biophys.J.}. Equivalence between queueing systems and transcription models has recently been systematically established by \cite{Solvingstochasticgeneexpression2024Biophys.J.}. Techniques from queueing theory can often circumvent CMEs and may extend to non-Markovian models. 
Unfortunately, these methods may suffer from limited physical interpretability and substantial computational cost. Less common approaches like the Poisson representation \cite{Poissonrepresentationbridge2023J.R.Soc.Interface} exist but are not analyzed in detail in this article. 

Furthermore, when working with CMEs, most previous studies consider only the steady-state solution by manually setting all time derivatives to zero. In such cases, the existence of equilibrium is presumed, which is not necessarily valid. Also, the dynamical behavior of the chemical reaction kinetics remains poorly characterized. Both the conditions guaranteeing convergence to equilibrium and the corresponding relaxation times are not well understood.

In this article, we develop a unified and comprehensive framework for studying stochastic transcription kinetics by analyzing a stochastic chemical reaction model of transcription with complicated gene-state architecture and transcription-induced state transitions. First, we show that the CME for this general model is exactly solvable. We directly present an explicit time-dependent solution to the corresponding CME by solving the partial differential equation system obtained via the standard generating function method. However, the resulting probability mass function of mRNA counts is typically algebraically unwieldy and thus impractical for numerical computation. Second, following the standard binomial moment method, we obtain time-dependent expressions for binomial moments of arbitrary order, and steady-state binomial moments then follow through taking limit with respect to time. Third, a sharp upper bound for binomial moments is derived by adapting an analogue of Theorem 4.1.2 in \cite{MatrixComputations2013} and using techniques from functional analysis. Subsequently, an upper bound for the probability mass function is also derived, which shows that the probability distribution of mRNA counts is always bounded by a constant multiple of a Poisson distribution. These two inequalities are later used to rigorously prove the main result of Heavy-Tailed Law \cite{ExtrinsicNoiseHeavyTailed2020Phys.Rev.Lett.,Exactlysolvablemodels2020J.Chem.Phys.}, stating that intrinsic noise alone never leads to a heavy-tailed distribution of mRNA counts. Explicit estimation for truncation error in practical implementation is also provided using these inequalities. To summarize, the main approach builds upon standard tools, handling the CME using the generating function method and the binomial moment method. The main novelties of our work include demonstrating the corresponding CME is exactly solvable, presenting dynamical behavior of binomial moments, and deriving inequalities with apparent physical meaning. In derivations, we introduce some non-trivial and original techniques, based on ideas from matrix analysis and functional analysis.

This article is organized as follows. We first present the method in detail. Complete derivations are provided in the Appendix. Then we validate the theoretical results using standard numerical methods, including SSA and FSP. Two important special cases are recovered from the general model: the Telegraph model and a Markovian model under renewal condition \cite{Solvingstochasticgeneexpression2024Biophys.J.}. Using mathematical analysis, we establish consistency with previous studies. Also, we briefly show how intrinsic and extrinsic noise can be studied in our model. Finally, we generalize the main conclusion of the Heavy-Tailed Law and provide a rigorous proof.

\section{Methods}\label{mainresults}
\subsection{A General Transcription Model}
In this article, we mainly study a fine-grained model of transcription with complicated gene-state architecture, as illustrated in the reaction scheme \eqref{Reaction}. We note that this general model \eqref{Reaction} was first proposed in \cite{Solvingstochasticgeneexpression2024Biophys.J.} (multi-step degradation of mRNA is not considered here), but a unified approach to its study remains absent in the literature. \eqref{Reaction} can be described either deterministically by the reaction rate equation \cite{StochasticProcessesPhysics2007} or stochastically by the CME. However, as previously noted, stochasticity is a central feature in gene expression dynamics, and the deterministic model fails to capture many phenomena. For instance, fluctuation in mRNA copy number and first passage time \cite{stochasticvsdeterministic2024Nat.Commun.} cannot be well described within a deterministic framework. 

\begin{equation}\label{Reaction}
\begin{aligned}
    &\ce{\mathcal{S}_i ->[$a_{i,j}$] \mathcal{S}_j}\;\;(i\neq j,1\leq i,j\leq N)\\
    &\ce{\mathcal{S}_i ->[$b_{i,j}$] \mathcal{S}_j + \textbf{mRNA}}\;\;(1\leq i,j\leq N)\\
    &\ce{\textbf{mRNA} ->[$\delta$] $\emptyset$}
\end{aligned}
\end{equation}

In \eqref{Reaction}, $\mathcal{S}_i\;(1\leq i\leq N)$ denote $N$ different gene states, where $N$ is referred to as the order of the model \eqref{Reaction} in this article, and $\textbf{mRNA}$ denotes an mRNA molecule. Reactions in the first row represent transitions among different states without transcription, while reactions in the second row represent transitions with production of a single mRNA molecule. The last reaction denotes hydrolysis of mRNA. In this model, arbitrarily many states of gene are allowed and transitions occur with or without production of mRNA molecules. If necessary, increase $N$ to ensure that each transition represents a single-step biochemical reaction. Note that reactants not directly involved in the above processes have been omitted to underscore the reaction structure. For instance, if translation of protein from mRNA and degradation of protein are considered \cite{Analyticaldistributionsstochastic2008Proc.Natl.Acad.Sci.U.S.A.}, the marginal distribution of mRNA molecules will coincide with the results presented in this article. Tuple $(\mathcal{S}(t),M(t))_{t\geq0}$ constitutes a continuous-time Markov chain with state space $\{1,2,\cdots,N\}\times\mathbb{N}$, where $\mathcal{S}(t)$ denotes the gene state and $M(t)$ denotes mRNA copy number at time $t$. Note that both $\mathcal{S}(t)$ and $M(t)$ are integer-valued random variables at any fixed time. 

We note that an arbitrary gene-state structure in \eqref{Reaction} is theoretically motivated and relatively flexible. In practice, designing an appropriate gene-state structure can be conceptually challenging.

\subsection{Chemical Master Equation and its Solution}
Let $a_{i,j}\;(1\leq i\neq j\leq N)$ denote the transition rate without transcription from state $\mathcal{S}_i$ to $\mathcal{S}_j$, and $b_{i,j}\;(1\leq i,j\leq N)$ the rate with transcription from state $\mathcal{S}_i$ to $\mathcal{S}_j$. Abbreviate $a_{i,i}:=-\sum_{\substack{k=1\\k\neq i}}^Na_{i,k}-\sum_{k=1}^Nb_{i,k}$ and $D_0:=(a_{i,j})_{N\times N}, D_1:=(b_{i,j})_{N\times N}$. Hydrolysis of mRNA molecules is assumed to be independent of the gene state, and has constant rate $\delta$. Let $\mathbb{P}_{i,j}(m;t)$ denote the probability that $m$ mRNA molecules are present in the system and the gene happens to be in state $\mathcal{S}_j$ at time $t$, given initial Dirac distribution $\mathcal{S}(0)=\mathcal{S}_i$ and $M(t)=0$. The corresponding CME is ($\forall$ $1\leq i,j\leq N$ and $m\in\mathbb{N}$)
\begin{equation}\label{CME}
\begin{aligned}
    \frac{\partial}{\partial t}\mathbb{P}_{i,j}(m;t)=&\sum_{\substack{k=1\\k\neq j}}^Na_{k,j}\mathbb{P}_{i,k}(m;t)+\sum_{k=1}^Nb_{k,j}\mathbb{P}_{i,k}(m-1;t)+(m+1)\delta\,\mathbb{P}_{i,j}(m+1;t)\\&-m\delta\,\mathbb{P}_{i,j}(m;t)
    -\sum_{\substack{k=1\\k\neq j}}^Na_{j,k}\mathbb{P}_{i,j}(m;t)-\sum_{k=1}^Nb_{j,k}\mathbb{P}_{i,j}(m;t).
\end{aligned}
\end{equation}
$\mathbb{P}_{i,j}(-1;t)$ is conventionally taken as null. Let $\mathbb{P}(m;t)\in\mathbb{R}^{N\times N}$ be a matrix whose $(i,j)$-th entry is $\mathbb{P}_{i,j}(m;t)$. Following standard generating function method, or equivalently, Laplace transform, define the matrix-form time-dependent generating function to be $\mathcal{G}(t,z)$, namely $\mathcal{G}(t,z):=\sum_{m=0}^\infty z^m\mathbb{P}(m;t)$. By rearranging, we convert the original chemical master equation \eqref{CME} into a partial differential equation system.
\begin{equation}\label{CME2}
\begin{aligned}
\frac{\partial}{\partial t}\mathcal{G}(t,z)=\mathcal{G}(t,z)D_0+z\,\mathcal{G}(t,z)D_1+(1-z)\,\delta\,\frac{\partial}{\partial z}\mathcal{G}(t,z).
\end{aligned}
\end{equation}
Inspired by the power series method for solving differential equations, we constructively present the analytical expression of $\mathcal{G}(t,z)$, the time-dependent solution to \eqref{CME2} under initial condition $\mathcal{G}(0,z)=\bm{I}_N$. Then the explicit expression of $\mathbb{P}_{i,j}(m;t)$ is obtained by reorganizing the generating function according to the power of $z$. When $m\geq 1$,
\begin{equation}\label{density}
\begin{aligned}
\mathbb P_{i,j}(m;t)&=\sum_{k=m}^{\infty} \int_{\Omega_k}
\sum_{\substack{I\subseteq\{1,\cdots,k\}\\ \# I=m}}
\Bigg[\prod_{s\in I}\alpha_s \prod_{r\in I^{\mathrm c}} (1-\alpha_r) \Bigg]
\bm{e}_i^\top K(t;t_1,\cdots,t_k)\bm{e}_j\mathrm{d}\bm{t};
\end{aligned}
\end{equation}
when $m=0$, we have
\begin{equation}\label{density0}
\begin{aligned}
\mathbb{P}_{i,j}(0;t)=&\bm{e}_i^\top\mathrm{e}^{D_0t}\bm{e}_j+\sum_{k=1}^\infty\int_{\Omega_k}\Bigg[\prod_{s=1}^{k}\bigl(1-\alpha_s\bigr)\Bigg]\bm{e}_i^\top
K(t;t_1,\cdots,t_k)\bm{e}_j\mathrm{d}\bm{t}.
\end{aligned}
\end{equation}
In \eqref{density} and \eqref{density0}, $\Omega_k:=\{(t_1,t_2,\cdots,t_k)\mid0\leq t_1\leq t_2\leq\cdots\leq t_k\leq t\}\subseteq\mathbb{R}^k$; $\#$ denotes the cardinality of a set; $\alpha_s:=\exp(-\delta t+\delta t_s)$; $I^c$ denotes the complement of $I$ with respect to $\{1,\cdots,k\}$; $\{\bm{e}_s\}_{s=1}^N$ are standard basis vectors in $\mathbb{R}^N$ where $\bm{e}_s$ is the column vector with a $1$ in the $s$-th position and $0$ elsewhere. The kernel function is defined as follows: 
$K(t;t_1):=\mathrm{e}^{D_0t_1}D_1\mathrm{e}^{D_0(t-t_1)}$, $K(t;t_1,t_2):=\mathrm{e}^{D_0t_1}D_1\mathrm{e}^{D_0(t_2-t_1)} D_1\mathrm{e}^{D_0(t-t_2)}$, and 
\newline
$K(t;t_1,\cdots,t_k):=\mathrm{e}^{D_0t_1}D_1\mathrm{e}^{D_0(t_2-t_1)}D_1\cdots D_1\mathrm{e}^{D_0(t_k-t_{k-1})} D_1\mathrm{e}^{D_0(t-t_k)}$ for $k\geq3$. The exponential of a square matrix, say $C$, is defined as $\mathrm{e}^C:=\sum_{k=0}^\infty C^k/k!$ (see Chapter $9$ of \cite{MatrixComputations2013}). We note that the summation over $I$ in \eqref{density} ranges over all unordered $m$-element subsets without repetition, and, when $I^c$ is the empty set, we take the corresponding term in the product as $1$. Detailed derivation is in Appendix, Section A. 

Although equations of the same form as \eqref{CME2} have been proposed in earlier studies for special cases relative to \eqref{Reaction} \cite{MultimodalityFlexibilityStochastic2013Bull.Math.Biol.,PromotermediatedTranscriptionalDynamics2014Biophys.J.}, it has never been found to be exactly solvable. Additionally, based on equivalence to queueing systems established by \cite{Solvingstochasticgeneexpression2024Biophys.J.}, an explicit expression of $\mathcal{G}(t,z)$ can actually be founded in existing results on the infinite-server queuing system with the Markovian arrival process \cite{AnalysisInfiniteServerQueue2002QueueingSyst.}, which provides a probabilistic interpretation of the expression of $\mathcal{G}(t,z)$. 
However, our derivation is based on CMEs and thus differs essentially from that in queueing theory. To the best of our knowledge, explicit expressions of the probability mass function \eqref{density} and \eqref{density0} are original.

Expressions \eqref{density} and \eqref{density0} reveal the solvability of CMEs in transcription models, explaining why analytical results of the probability mass function can be expected in some special cases \cite{Exactdistributionsstochastic2022Phys.Rev.E,Exactlysolvablemodels2020J.Chem.Phys.,AnalyticalTimeDependentDistributions2023SIAMJ.Appl.Math.,Analyticalresultsmultistate2012SIAMJ.Appl.Math.}. Nevertheless, though theoretically satisfactory, \eqref{density} and \eqref{density0} are too complicated for numerical computation.

\begin{figure*}[h!]
    \centering
    \includegraphics[width=\textwidth]{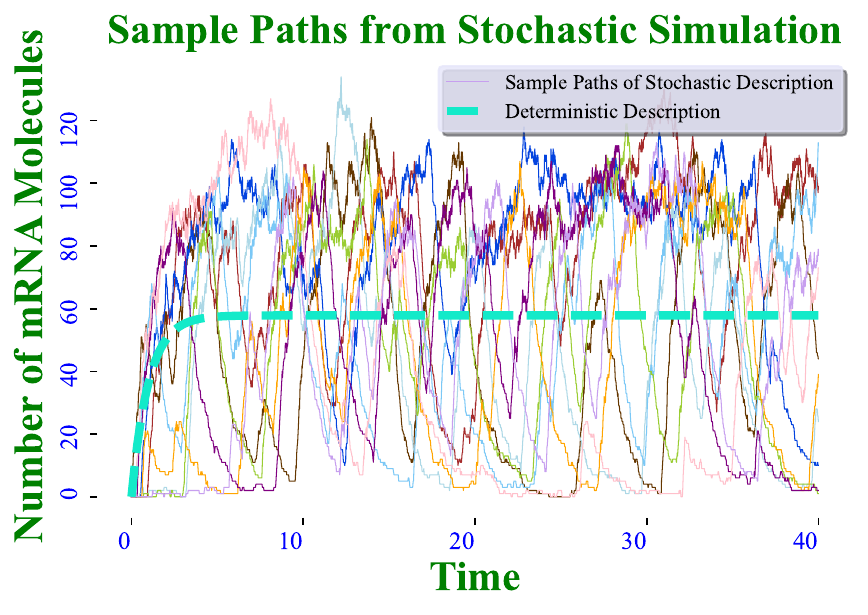}
    \caption{\textbf{Trajectories from Stochastic Simulation}: 10 sample paths of stochastic simulation of the reaction system \eqref{Reaction} are plotted in the above illustration. The parameters are set as $D_0=$\scalebox{0.5}{$\begin{pmatrix}
-2.11 &  0.1  &  0.0  &  0.0  &  0.01 \\
  0.1  & -2.61 &  0.5  &  0.0  &  0.01 \\
  0.1  &  0.2  & -5.4  &  0.0  &  0.1  \\
  0.1  &  0.1  &  0.3  & -3.5  &  0.0  \\
  0.1  &  0.1  &  0.1  &  0.1  & -100.4
\end{pmatrix}$}, $D_1=$\scalebox{0.5}{$ \begin{pmatrix}
0 & 0 & 1 & 0 & 1  \\
 1 & 1 & 0 & 0 & 0  \\
 1 & 2 & 1 & 1 & 0  \\
 1 & 0 & 0 & 1 & 1  \\
 0 & 0 & 0 & 0 & 100
\end{pmatrix}$} and $\delta=1$. The initial state of the system is $M(0)=0$ and $\mathcal{S}(0)=\mathcal{S}_1$. The bold dashed line of \textit{deterministic description} is the time-dependent solution of the reaction rate equation describing \eqref{Reaction} under given initial condition (assume the underlying Markov chain characterized by $D$ is in equilibrium). We sample $4000$ points equally spaced in the time interval $[0,40]$. Python package \texttt{GillesPy2} \cite{GillesPy2BiochemicalModeling2023Lett.Biomath.} is used.}
    \label{SSAtraj1}
\end{figure*}

In \autoref{SSAtraj1}, $10$ sample paths generated with stochastic simulation, and the time-dependent solution of the reaction rate equation describing \eqref{Reaction} are plotted. The sample paths are observed to fluctuate remarkably around the deterministic solution, and stationary point of the reaction rate equation is actually rarely achieved by sample paths. In this example, the steady-state distribution of mRNA counts is concentrated both far above and far below the stationary point of the reaction rate equation.

\subsection{Binomial Moment Method}
For mathematical and computational tractability, we instead consider binomial moments of mRNA counts, following the standard binomial moment approach. 

We begin by introducing the widely adopted binomial moment method, now considered one standard technique in the CME studies. The binomial moment method has actually been implicitly used in queueing theory for a long time \cite{$GI^Xinfty$system1990J.Appl.Probab.,AnalysisInfiniteServerQueue2002QueueingSyst.,$Minfty$queue1986J.Appl.Probab.}, and is systematically proposed in the context of chemical reaction kinetics by \cite{BinomialMomentEquations2011Phys.Rev.Lett.,Stochasticanalysiscomplex2012Phys.Rev.Ea}. The moment-convergence method \cite{momentconvergencemethodstochastic2016J.Chem.Phys.a} is one useful generalization of the binomial moment method. Recall that, for a given random variable taking values in nonnegative integers, say $X$, its binomial moments, denoted by $\{B_m\}_{m\in\mathbb{N}}$, are defined formally as the Taylor coefficients of its generating function expanded at $z=1$. One can easily prove that, given the probability mass function denoted by $\{\mathcal{P}_n\}_{n\in\mathbb{N}}$, we have (should the binomial moments exist) \cite{momentconvergencemethodstochastic2016J.Chem.Phys.a}
\begin{equation}\label{distribution_reverse}
    B_m=\sum^{\infty}_{n=m}\binom{n}{m}\mathcal{P}_n,\; m\in\mathbb{N},
\end{equation}
where $\binom{m}{n}$ is the combinatorial coefficient. Therefore, binomial moments are actually linear combinations of moments \cite{CourseProbabilityTheory2000} defined as $\langle X^m\rangle:=\sum^{\infty}_{n=0}n^m\mathcal{P}_n,\;m\in\mathbb{N}$, and thus binomial moments of arbitrary order exist if and only if moments of arbitrary order all exist. The binomial moment method is actually studying a given random variable by investigating its moments instead of the probability mass function. 
In many cases, differential equations describing binomial moments are more feasible to solve and easier to truncate than CMEs \cite{BinomialMomentEquations2011Phys.Rev.Lett.,Stochasticanalysiscomplex2012Phys.Rev.Ea,momentconvergencemethodstochastic2016J.Chem.Phys.a}. This is mainly because binomial moments, or moments, represent macroscopic physical observables of the distribution, such as expectation and variance. Also, the implicit constraint $\sum_{n=0}^\infty \mathcal{P}_n=1$ in CMEs is explicitly captured by $B_0=1$ or $\langle X^0\rangle=1$. Most importantly, the probability mass function can also be easily reconstructed by binomial moments according to the identity \cite{momentconvergencemethodstochastic2016J.Chem.Phys.a}
\begin{equation}\label{distribution}
   \mathcal{P}_n=\sum^{\infty}_{m=n}\left(-1\right)^{m-n}\binom{m}{n}B_m,\; n\in\mathbb{N}.
\end{equation}
Nevertheless, we emphasize that the convergence of \eqref{distribution} requires careful verification, since moments may not uniquely determine the underlying probability distribution. This is essentially the well-known moment problem \cite{CourseProbabilityTheory2000} in probability.

Note that $D:=D_0+D_1\in\mathbb{R}^{N\times N}$ is a $Q$-matrix, meaning off-diagonal entries are nonnegative and row sum is null. Assume $D$ to be irreducible throughout the article. This assumption is physically reasonable, since most biochemical processes are reversible. For simplicity, we assume the underlying Markov chain of gene states (characterized by $D$) has converged to equilibrium, and take the unique invariant distribution to be $\bm{\pi}\in\mathbb{R}^{N\times 1}$. By definition, $\bm{\pi}$ is the unique solution to $\bm{\pi}^\top D=\bm{0}_{1\times N}$ and $\bm{\pi}^\top \bm{1}=1$, where $\bm{1}$ is a $N\times 1$ column vector of all ones. This assumption is reasonable if the architecture of gene states reconstruct substantially slow compared with the transcription kinetics. We now cease tracking microstates of the gene and focus on the coarse-grained number of mRNA molecules. By summing over current states and setting the initial distribution to be $\bm{\pi}$, we define the steady-state, scalar binomial moments and probability mass function of mRNA counts ($m,n\in\mathbb{N}$) $B_m:= \frac{1}{m!}\lim_{t\rightarrow\infty}[\frac{\partial^m}{\partial z^m}\bm{\pi}^\top\mathcal{G}(t,z)\bm{1}\big|_{z=1}]$ and $\mathcal{P}_n:= \frac{1}{n!}\lim_{t\rightarrow\infty}[\frac{\partial^n}{\partial z^n}\bm{\pi}^\top\mathcal{G}(t,z)\bm{1}\big|_{z=0}]$
should they exist. We assert that steady-state binomial moments $\{B_m\}_{m\geq 1}$ not only exist but have the following compact expressions
\begin{equation}\label{binomial}
\begin{aligned}
B_1&=\frac{1}{\delta}\bm{\pi}^\top D_1\bm{1},\\
B_m&=\frac{1}{m\delta}\bm{\pi}^\top D_1\left(\delta\bm{I}_N-D\right)^{-1}D_1\left(2\delta\bm{I}_N-D\right)^{-1}\cdots D_1
   \left[(m-1)\delta\bm{I}_N-D\right]^{-1}D_1\bm{1},\; m\geq 2,
\end{aligned}
\end{equation}
where $\bm{I}_N\in\mathbb{R}^{N\times N}$ is the identity matrix. Note that $\lambda\bm{I}_N-D$ is a strictly diagonally dominant matrix for $\lambda>0$, thus nonsingular by L\'{e}vy-Desplanques theorem \cite{MatrixAnalysis2012}. The derivation of \eqref{binomial} is in the Appendix, Section B. 

We acknowledge that the expression in the same form as \eqref{binomial} already exists, dating back to results in queueing theory \cite{$Minfty$queue1986J.Appl.Probab.} and also appearing in studies of transcription models with multiple gene states \cite{MultimodalityFlexibilityStochastic2013Bull.Math.Biol.,PromotermediatedTranscriptionalDynamics2014Biophys.J.}. These previous studies can all be included in the special case where $D_1$ is diagonal, compared with \eqref{Reaction}. Nevertheless, the essential difference between our article and the literature lies in the approach we derive \eqref{binomial}. Most previous studies, including \cite{MultimodalityFlexibilityStochastic2013Bull.Math.Biol.,$Minfty$queue1986J.Appl.Probab.,PromotermediatedTranscriptionalDynamics2014Biophys.J.}, obtain \eqref{binomial} by formally deriving the binomial moment equations and considering only the fixed point of this ordinary differential hierarchy. For first-order chemical reaction systems like \eqref{Reaction}, binomial moment equations are closed, in the sense that the evolution of one binomial moment is independent of higher-order binomial moments. Therefore, fixed point of this hierarchy can be easily derived by recurrence and no calculus is involved. 
However, we derive \eqref{binomial} via taking limit with respect to time in the analytical expression of $B_m(t):=\frac{\partial^m}{\partial z^m}\bm{\pi}^\top \mathcal{G}(t,z)\bm{1}\big|_{z=1}$ (see the Appendix, Section B). In this sense, we additionally establish the stability of this fixed point. 

The recurrence method proposed in \cite{Exactlysolvablemodels2020J.Chem.Phys.} also implicitly exploits \eqref{binomial}, but it is the first time the numerical advantage of \eqref{binomial} is highlighted. We will provide further numerical analysis in the subsection Numerical Analysis.

\begin{figure*}[b!]
    \centering
    \includegraphics[width=0.8\textwidth]{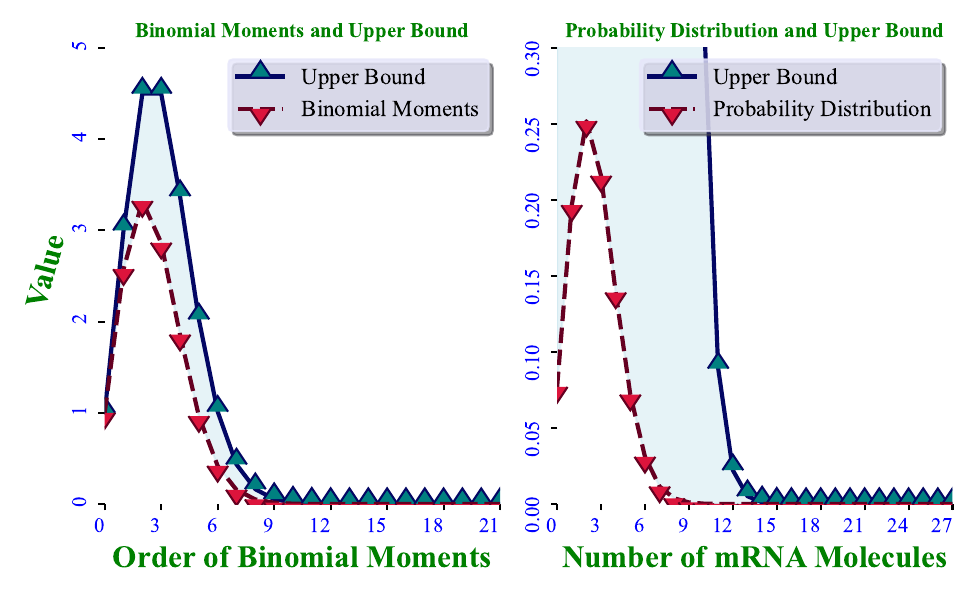}
    \caption{\textbf{Upper Bound for Binomial Moments and Probability Distribution of mRNA molecules}: In the left panel, binomial moments and the upper bound are computed according to \eqref{binomial} and \eqref{converge}, respectively. In the right panel, the upper bound of the probability mass function is given by \eqref{Bound1}. Without loss of generality, we normalize all the parameters by dividing $\delta$. In the above illustrations, parameters are both set as $D_0 =$\scalebox{0.5}{$\begin{pmatrix}-7 & 2 & 3 & 0 \\3 & -12 & 1 & 5 \\1 & 2 & -12 & 6 \\8 & 8 & 0 & -19    \end{pmatrix}$} and $D_1=$\scalebox{0.5}{$\begin{pmatrix}
    1 & 1 & 0 & 0 \\1 & 0 & 1 & 1 \\0 & 1 & 1 & 1 \\1 & 1 & 0 & 1\end{pmatrix}$}.}
    \label{Bound}
\end{figure*}

\subsection{Two Inequalities}
Furthermore, we estimate the asymptotic behavior of $\{B_m\}_{m\in\mathbb{N}}$ and $\{\mathcal{P}_n\}_{n\in\mathbb{N}}$. Note that $B_0$ is always unity by definition. For a strictly row diagonally dominant matrix $C=(c_{i,j})_{N\times N}$ with $\Theta$ defined as $\min_{1\leq i\leq N}\left(\mid c_{i,i}\mid-\sum_{j\neq i}\mid c_{i,j}\mid\right)$, $\lVert C^{-1}\rVert_\infty$ is bounded by $\Theta^{-1}$ from above (an analogue of Theorem 4.1.2 in \cite{MatrixComputations2013}). Recall that infinity norm of a matrix, denoted by $\lVert \cdot\rVert_\infty$, is the maximum absolute row sum of this matrix. The following result confirms convergence of $\{B_m\}_{m\in\mathbb{N}}$ and further delineates the velocity of convergence. To specify, 
\begin{equation}\label{converge}
    B_m\leq\frac{1}{m!}\left(\frac{\lVert D_1\rVert_{\infty}}{\delta}\right)^m.
\end{equation}
In particular, $\{B_m\}_{m\in\mathbb{N}}$ converges to zero. See the left panel in \autoref{Bound}. Deriving \eqref{converge} follows an approach similar to bounding the operator norm of a bounded linear functional, and we borrow ideas from functional analysis. The details can be found in the Appendix, Section C. We can also prove that exact probability mass function of mRNA molecules $\{\mathcal{P}_n\}_{n\in\mathbb{N}}$ is uniquely determined by binomial moments $\{B_m\}_{m\in\mathbb{N}}$ and can be given according to \eqref{distribution}. Substituting \eqref{binomial} into \eqref{distribution}, a theoretically exact and computationally feasible approach to obtaining steady-state distribution of mRNA copy number is established. Note that, by proving the convergence of series \eqref{distribution}, we rigorously establish the existence and uniqueness of steady-state mRNA distribution given arbitrary parameters in \eqref{Reaction}. Note that the convergence of \eqref{distribution} needs to be rigorously proved, since, in general, moments do not uniquely determine a probability distribution. 

To quantify the asymptotic behavior of $\{\mathcal{P}_n\}_{n\in\mathbb{N}}$, we have
\begin{equation}\label{Bound1}
\begin{aligned}
   \mathcal{P}_n\leq \frac{1}{n!}\mathrm{e}^{\lVert D_1\rVert_{\infty}/\delta}\left(\frac{\lVert D_1\rVert_{\infty}}{\delta}\right)^n, \quad n\in\mathbb{N}.
\end{aligned}
\end{equation}
This inequality demonstrates that the probability mass function of mRNA copy number is always bounded from above by a constant multiple of the mass function of the Poisson distribution with parameter $\lVert D_1\rVert_{\infty}/\delta$. To emphasize the usefulness of this inequality, we utilize \eqref{Bound1} to generalize and rigorously prove the main conclusion of the Heavy-Tailed Law in the last subsection of RESULTS. Since the upper bound in \eqref{Bound1} converges to zero as $n$ grows, sequence $\{\mathcal{P}_n\}_{n\in\mathbb{N}}$ tends to null. Note also that this bound becomes practical only when $n$ is relatively large because $\mathcal{P}_n\leq1$ always holds. See the right illustration in \autoref{Bound}. Proof of \eqref{Bound1} is in the Appendix, Section D.

Conventionally, when analyzing chemical reaction systems like \eqref{Reaction}, previous studies typically assume in advance the existence of equilibrium, and calculate the steady-state solution with all time derivatives in CMEs or binomial moment equations set to zero. In this article, instead, we construct directly an analytical formulation of the time-dependent solution and then take the limit with respect to time. Thereby, our results additionally illustrates that the reaction system \eqref{Reaction} admits a unique steady state and always converges to equilibrium after a sufficiently long period of time. The rate of approaching the equilibrium may also be estimated through further analysis built upon the time-dependent solution \eqref{density} and \eqref{density0}.

\section{Results}
This section is structured as follows. First, we verify the analytical results presented above with numerical methods including SSA and FSP. In the following two subsections, we investigate two special cases of the general model, respectively. Then, we analyze the implementation of our result from a numerical perspective. Finally, we briefly discuss the fluctuation of mRNA copy number in transcription models.

\subsection{Verification via Stochastic Simulation and Finite State Projection}
\begin{figure*}[t!]
    \centering
    \includegraphics[width=0.8\textwidth]{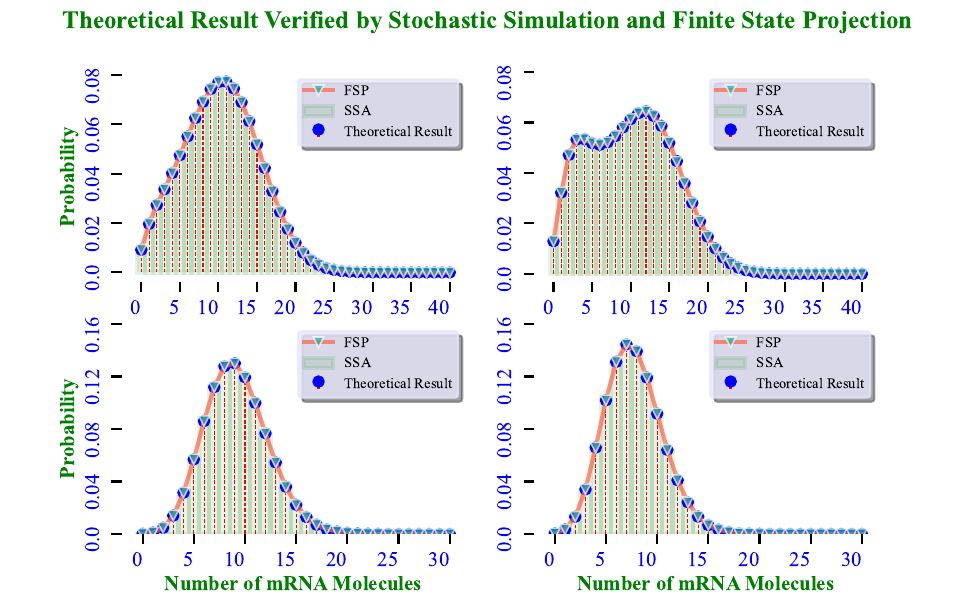}
    \caption{\textbf{Probability Distribution of mRNA Counts through SSA, FSP, and Our Theoretical Results}: Steady-state distributions of mRNA counts in four examples are computed using three different approaches. The histograms are each generated from $1\times10^5$ trajectories using SSA, truncated at dimensionless time $t=40$. \texttt{GillesPy2} is implemented with C++ solver. The line plots are generated using FSP, where the truncation is determined such that truncation error is below $1\times10^{-5}$. The steam plots are generated according to analytical results \eqref{binomial} and \eqref{distribution}. For parameters in these four model, refer to Jupyter Notebook \texttt{ParametersFig3.ipynb}. These four models are designed to represent progressively increasing complexity, specifically of orders $3$, $5$, $10$, and $20$.}
    \label{SSAFSP}
\end{figure*}

SSA and FSP are widely recognized as classical and general-purpose numerical approaches for studying stochastic chemical reaction kinetics. We now validate our analytical results, \eqref{binomial} and \eqref{distribution}, through stochastic simulations of the reaction system \eqref{Reaction}, and implementing FSP. As shown in \autoref{SSAFSP}, the analytical steady-state distribution of mRNA molecules is in excellent agreement with results obtained from both SSA and FSP.

\subsection{Special Case I: The Telegraph Model}
Recall that in the Telegraph model, the gene switches memorylessly between two states and the production of mRNA molecules occurs only when the gene is in the active state. 
With $D_0=
 \begin{pmatrix}
-k_1-\lambda & k_1 \\
k_2 & -k_2
\end{pmatrix}$ and $D_1=\text{diag}[\lambda,0]$, our model \eqref{Reaction} reduces to the Telegraph model. Since the inverse of a second-order matrix can be expressed explicitly, further reduction in \eqref{binomial} becomes possible in this simple model. It follows that 
\begin{equation}\label{telegraph}
\begin{aligned}
    B_m&=\frac{1}{m!}\left(\frac{\lambda}{\delta}\right)^m\frac{(k_2/\delta)^{(m)}}{(k_1/\delta+k_2/\delta)^{(m)}},\;m\in\mathbb{N},
\end{aligned}
\end{equation}
where $x^{(m)}:=\Gamma(x+m)/\Gamma(x)$ is the Pochhammer symbol. $\Gamma(x)$ is Gamma function. By substituting \eqref{telegraph} into \eqref{distribution}, it can be derived that
\begin{equation}\label{1F1s}
\begin{aligned}
    \mathcal{P}_n&=\frac{1}{n!}\frac{\lambda^n(k_2/\delta)^{(n)}}{\delta^n(k_1/\delta+k_2/\delta)^{(n)}}\;\mathllap{_{1}}F_1\left(\frac{k_2}{\delta}+n,\frac{k_1+k_2}{\delta}+n;-\frac{\lambda}{\delta}\right),\;n\in\mathbb{N},
\end{aligned}
\end{equation}
where \;$\mathllap{_{1}}F_1(a,b;z):=\sum_{n=0}^\infty\frac{a^{(n)}}{b^{(n)}}\frac{z^n}{n!}$ denotes the confluent hypergeometric function. \eqref{1F1s} is consistent with previous results \cite{Exactlysolvablemodels2020J.Chem.Phys.,MarkovianModelingGeneProduct1995Theor.Popul.Biol.}. A detailed calculation can be found in the Appendix, Section F.

Although formulation of $\{\mathcal{P}_n\}_{n\in\mathbb{N}}$ in \eqref{1F1s} seems formally compact, the confluent hypergeometric function is, in essence, an abbreviation of an infinite series. Evaluating the value of confluent hypergeometric function at a given point possesses approximately the same computational complexity as one calculation in \eqref{distribution}, and it involves truncation error as well. From this standpoint, \eqref{1F1s} is similar, theoretically, to reconstructing $\{\mathcal{P}_n\}_{n\in\mathbb{N}}$ from binomial moments according to \eqref{distribution}. This argument also holds for other well-known results involving the generalized hypergeometric function \cite{Exactdistributionsstochastic2022Phys.Rev.E,Exactlysolvablemodels2020J.Chem.Phys.,AnalyticalTimeDependentDistributions2023SIAMJ.Appl.Math.,Solvingstochasticgeneexpression2024Biophys.J.}. However, special function modules, such as \texttt{scipy.special} in Python, may have built-in implementation of the confluent hypergeometric function, where the numerical computation is carefully designed to mitigate instability. In this case, expressions in the form of the confluent hypergeometric function or its variants are preferable, because numerical computation based on \eqref{binomial} and \eqref{distribution} can be highly unstable if poorly implemented. We will discuss numerical challenges in detail in the section Numerical Analysis.

\subsection{Special Case II: Markovian Models Under Renewal Condition}
Recall that a renewal process is a counting process whose interarrival times are independent and identically distributed according to a given distribution with probability density function $f(t)$. Note that $f(t)$ is assumed to be integrable in most cases. Previous studies predominantly focus on the case where production of mRNA molecules becomes a renewal process, and queueing theory is introduced to obtain powerful results in a relatively general framework \cite{IntrinsicNoiseStochastic2011Phys.Rev.Lett.,Inferringtranscriptionalbursting2023R.Soc.OpenSci.,QueuingModelsGene2020Biophys.J.,Solvingstochasticgeneexpression2024Biophys.J.}. In \eqref{Reaction}, production of mRNA molecules becomes a renewal process if and only if the following renewal condition \cite{Solvingstochasticgeneexpression2024Biophys.J.} holds: When $\sum_{j=1}^N b_{i,j}\neq 0$, $\frac{b_{i,j}}{\sum_{j=1}^N b_{i,j}}$ is independent of $i$.
Assuming $D_1\neq \bm{0}_{N\times N}$, this renewal condition is equivalent to $\text{rank}\,D_1=1$. 

We specifically term \eqref{Reaction} as a Markovian model under renewal condition when the aforementioned renewal condition is satisfied.
Note that the Telegraph model discussed in the previous subsection can also be incorporated into this framework; however, we treat it separately because its results permit further simplification. In general, the Markovian model under renewal condition can be interpreted alternatively as a queueing system $GI/M/\infty$, with probability density function of interarrival times being \cite{Solvingstochasticgeneexpression2024Biophys.J.}
\begin{align}\label{pdf}
    f(t)=\bm{\kappa}^\top\mathrm{e}^{D_0t}D_1\bm{1},\;t>0,
\end{align}
where $\bm{\kappa}\equiv(\kappa_j)_{1\leq j\leq N}:=(\frac{b_{i,j}}{\sum_{k=1}^N b_{i,k}})_{1\leq j\leq N}\in\mathbb{R}^{N\times 1}$ as long as $\sum_{k=1}^N b_{i,k}\neq 0$.

Queueing systems with arrivals following a renewal process have been thoroughly studied \cite{$GI^Xinfty$system1990J.Appl.Probab.}. The tractable nature of the renewal process yields satisfactory theoretical results. For the queueing system $GI/M/\infty$, the stationary binomial moments are given by \cite{$GI^Xinfty$system1990J.Appl.Probab.}
\begin{align}\label{GIqueue}
    B_m=\frac{1}{\alpha m\delta}\prod_{i=1}^{m-1}\frac{\mathcal{L}[f](i\delta)}{1-\mathcal{L}[f](i\delta)}, \; m\geq 2;\,B_1=\frac{1}{\alpha\delta},
\end{align}
where $\mathcal{L}[f](s):=\int_0^\infty \mathrm{e}^{-st}f(t)\mathrm{d}t,\;s>0$ is the Laplace transform of $f(t)$ and $\alpha:=\int_0^\infty tf(t)\mathrm{d}t$ is the expectation of inter-arrival time. 

Assuming the nonsingularity of $D_0$, binomial moments given in \eqref{GIqueue} can be reduced to \eqref{binomial}. The details can be found in the Appendix, Section G. We note that, by definition of the renewal process, holding time should be integrable, which is equivalent to the convergence of $\alpha$, or the nonsingularity of $D_0$. Actually, under the assumption $D$ is irreducible and $\text{rank}\,D_1=1$, $D_0$ is nonsingular. The rigorous proof is presented in the Appendix, which is non-trivial and can be seen as an analogue of Taussky theorem \cite{MatrixAnalysis2012} in matrix analysis.

\subsection{Numerical Analysis}
This subsection presents a systematic analysis of our theoretical results from a numerical perspective. We highlight their main advantage, namely the low computational complexity, and compare them with three widely-used approaches. Additionally, we analyze numerical errors in practical implementation, including truncation and floating point errors, and then propose strategies for their control, respectively.

\begin{figure*}[t!]
    \centering
    \includegraphics[width=\textwidth]{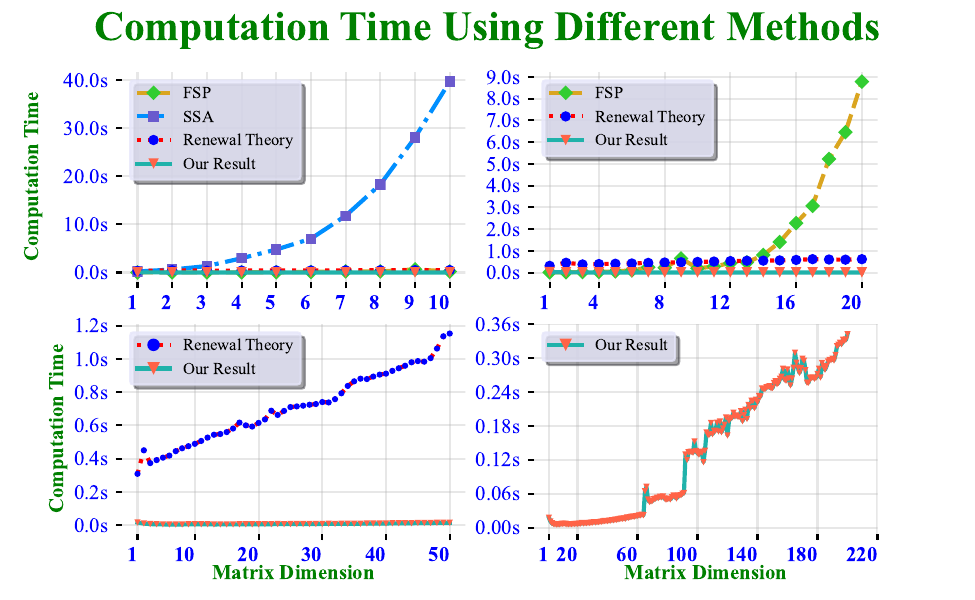}
    \caption{\textbf{Computation Time Using Different Methods}: Computation times are evaluated for a sequence of increasingly complicated models using four different approaches. Parameters are given in the form of $a_{i,j}=0(i\neq j)$ and $b_{i,j}=i$. As the dimension of $D_0$ or $D_1$ increases, the number of reaction pathways grows quadratically, resulting in increased model scale. SSA is performed for $1000$ trajectories and truncated at dimensionless time $t=10$ (\texttt{GillesPy2} is implemented with Python solver); FSP is implemented such that approximation error caused by truncating the CME is below $1\times10^{-5}$; Curves of \textit{Renewal Theory} and \textit{Our Result} are computation times of the first $200$ binomial moments (starting from $B_1$) based on \eqref{GIqueue} and \eqref{binomial}, respectively.
    In the top left panel, four different approaches (SSA; FSP; \textit{Renewal Theory}; \textit{Our Result}) are implemented to models with dimension $k\;\;(1\leq k\leq 10)$; in the top right panel, three approaches (FSP; \textit{Renewal Theory}; \textit{Our Result}) are implemented to models with dimension $k\;\;(1\leq k\leq20)$; in the bottom left panel, two approaches (\textit{Renewal Theory}; \textit{Our Result}) are implemented to models with dimension $k\;\;(1\leq k\leq50)$; in the bottom right panel, only \textit{Our Result} is implemented to models with dimension $k\;\;(1\leq k\leq200)$. To mitigate randomness, running times are all evaluated for three replicate computations and then averaged.}
    \label{Time}
\end{figure*}

From a numerical implementation perspective, the key advantage of \eqref{binomial} is its low computational complexity. As demonstrated in \autoref{Time}, our result \eqref{binomial} exhibits much lower computation time compared to three other methods. Actually, running times of SSA and FSP are both greatly understated in \autoref{Time}. For example, SSA is performed for only $1\times10^3$ trajectories and truncated at dimensionless time $t=10$, but in practice over $1\times10^5$ simulations may be needed and the simulation time $t=10$ is not long enough. Also, in \autoref{Time}, we only measure one final execution computing the matrix exponential $\exp{(t_0A_J)}$ (in fact, we directly compute its product with a given vector), where $A_J$ denotes the truncated coefficient matrix for the the original CME \eqref{CME} and $t_0$ represents the sampling time. The preceding iterative process of expanding the state subset, or equivalently the matrix $A_J$, to achieve a given error bound is excluded from the timing. However, separation of time scales can still be explicitly observed among four different approaches, as illustrated in \autoref{Time}. Specifically, general approaches including SSA and FSP quickly become infeasible for large-scale systems, while analytical results like \eqref{GIqueue} and \eqref{binomial} perform much better. Moreover, the computational complexity of \eqref{binomial} is significantly lower than that of \eqref{GIqueue}, because \eqref{binomial} involves only matrix inversion, matrix-matrix multiplication and matrix-vector multiplication, whereas \eqref{GIqueue} additionally involves numerical integration and matrix exponential, which are both computationally expensive. In particular, since $k\delta\bm{I}_N-D$ is diagonally dominant, we can directly perform LU factorization without pivoting (Theorem $4.1.1$ in \cite{MatrixComputations2013}). It involves $\frac{8}{3}N^3$ flops to calculate $\left(k\delta\bm{I}_N-D\right)^{-1}$ for given $k$ and $\frac{20}{3}(m-1)N^3$ flops to calculate $B_m$ in general. Recursive structure in \eqref{binomial} can also be utilized to reduce overall complexity. Note that our result remains computationally feasible for extremely high-dimensional models while other approaches generally fail.

Now we analyze computational errors arising from the numerical implementation of \eqref{binomial} and \eqref{distribution}. The overall errors consist of truncation error and floating point error. 

On one hand, as an infinite series, evaluating the value of $\mathcal{P}_n$ based on \eqref{distribution} inevitably introduces truncation error. Recall that we can prove series \eqref{distribution} converges using \eqref{converge}. Therefore, truncation error can be arbitrarily small if truncation point is chosen large enough. In particular, a priori estimation of truncation error is expected and is substantially favored over empirical choices. Nonetheless, except for FSP, few existing approaches to analyzing stochastic gene expression models effectively achieve this objective. In contrast, in our framework, we now give an explicit upper bound for truncation error, formulated below. Let $\widehat{\mathcal{P}_n}$ denote the computed value given the truncation point $M$. Assume $M>\max\{2n+2,8\mathrm{e}\left(\lVert D_1\rVert_{\infty}/\delta\right)^2\}$. Then 
\begin{equation}\label{bound2}
    \mid \mathcal{P}_n-\widehat{\mathcal{P}_n}\mid\leq \frac{1}{\sqrt{2\pi \mathrm{e}}n!2^{M-1}},
\end{equation}
which is proved in the Appendix, Section E. Thereby, truncation error can be well controlled using this inequality.

On the other hand, floating point error is introduced in both \eqref{binomial} and \eqref{distribution}. Computing binomial moments via \eqref{binomial} is relatively accurate and stable, because inversion of diagonally dominant matrix, matrix-matrix multiplication and matrix-vector multiplication are all controllable operations \cite{MatrixComputations2013}. However, floating point error can become extremely severe when reconstructing probability mass function based on \eqref{distribution}. Since binomial moments may rapidly drop below the threshold of machine precision while the combinatorial coefficients can grow very large, multiplying a small binomial moment by a large combinatorial coefficient suffers from precision loss. Thereby, particular care must be taken in practical implementations. The essentially same problem has been observed and carefully analyzed in \cite{Exactlysolvablemodels2020J.Chem.Phys.}. In the Jupyter Notebooks provided in the Supplementary Material, we compute terms $\binom{m}{n}B_m$ in \eqref{distribution} by first performing logarithmic-scale operations (use built-in function \texttt{math.lgamma(n+1)} to compute $\ln{(n!)}$\;) and then exponentiating the output. Our experience shows that this approach is much more stable than direct computation using \texttt{math.comb}.

Another possible way to avoid the above floating point error is to transform the probability mass function into binomial moments and conduct subsequent analysis completely based on binomial moments. Recall that given probability mass function, binomial moments can be obtained via \eqref{distribution_reverse}.

\subsection{Fluctuation of mRNA Copy Number in Transcription Models}\label{HTL}
Noise in gene expression within a supposedly identical ensemble arises from multiple sources \cite{UsingGeneExpression2012Science,NoiseGeneExpression2005Science}. Quantitative understanding of generation and propagation of noise is one of the primary objectives of modeling stochastic gene expression, and noise in transcription can predominantly contribute to protein-level fluctuation \cite{Regulationnoiseexpression2002NatureGenet.}. Experimentally, noise can be categorized into intrinsic and extrinsic noise. Intrinsic noise is the fluctuation originating from inherent stochasticity of biochemical reactions. The model \eqref{Reaction} and the results established above provide a general framework of investigating origins and consequences of intrinsic noise in transcription. Based on \eqref{binomial}, Fano factor, the ratio of variance to mean of mRNA copy number, is given by
\begin{align}\label{fano}
    FF=\frac{\bm{\pi}^\top D_1(\delta\bm{I}_N-D)^{-1}D_1\bm{1}}{\bm{\pi}^\top D_1\bm{1}}+1-\frac{\bm{\pi}^\top D_1\bm{1}}{\delta}.
\end{align}
Intrinsic noise, defined as the ratio of variance to mean square, is
\begin{align}\label{cv}
    \eta^2=\frac{\delta\bm{\pi}^\top D_1(\delta\bm{I}_N-D)^{-1}D_1\bm{1}}{(\bm{\pi}^\top D_1\bm{1})^2}+\frac{\delta}{\bm{\pi}^\top D_1\bm{1}}-1.
\end{align}
For specific gene-state structure where $(\delta\bm{I}_N-D)^{-1}$ has explicit expression, further reduction of Fano factor and intrinsic noise may be possible. Consistency of \eqref{fano} and \eqref{cv} with previous results can be verified \cite{Exactdistributionsstochastic2022Phys.Rev.E,Solvingstochasticgeneexpression2024Biophys.J.}. Deriving Fano factor and intrinsic noise from binomial moments is standard and details can be found in the Appendix, Section I. As illustrated in \autoref{noiseandfano}, intrinsic noise converges to zero as model size increases, implying that stochastic kinetics of mRNA molecules tends towards the deterministic description in these examples. Additionally, Fano factor is greater than $1$ except for the trivial $n=1$ case according to \autoref{noiseandfano}, meaning that distribution of mRNA copy number is super-Poissonian in these examples. 

\begin{figure*}[t!]
    \centering
    \includegraphics[width=0.7\textwidth]{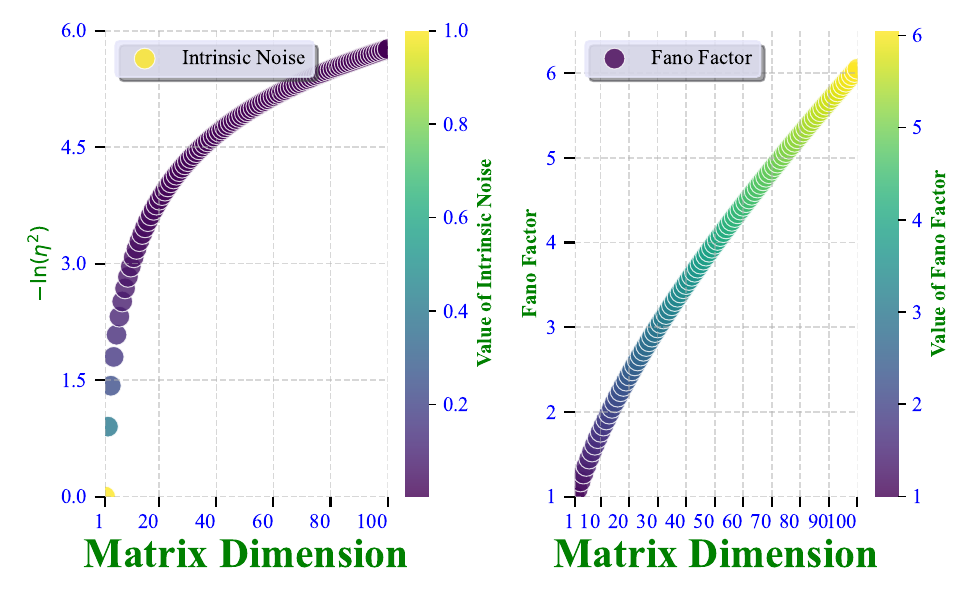}
    
    \caption{\textbf{Intrinsic Noise and Fano Factor}: Intrinsic noise and Fano factor are numerically evaluated for models of order $n$($1\leq n\leq 100$), with parameters the same as \autoref{Time}. In the left panel, intrinsic noise $\eta^2$ is computed based on \eqref{cv} and $-\ln(\eta^2)$ is used to generate the illustration; in the right panel, Fano factor is computed based on \eqref{fano}.}
    \label{noiseandfano}
\end{figure*}

The extrinsic noise refers to the fluctuation caused by environmental differences among cells, and can be further classified as local and global extrinsic noise \cite{NoiseGeneExpression2005Science}. Although the intrinsic noise can be readily included in some mathematical frameworks, such as CMEs, incorporating extrinsic noise is conventionally considered difficult. One insightful approach, proposed by \cite{ExtrinsicNoiseHeavyTailed2020Phys.Rev.Lett.}, is to utilize the parametric nature of the model. Parameters, which are $D_0$, $D_1$ and $\delta$ in our case, are assumed to follow a given probability distribution with density function $\rho(\bm{\theta})$, representing effects of the extrinsic perturbation. Since binomial moments have concise expressions given by \eqref{binomial}, posing the perturbation on binomial moments and then reconstructing the probability distribution is more tractable than posing perturbation directly on the probability mass function as in \cite{ExtrinsicNoiseHeavyTailed2020Phys.Rev.Lett.}. Thereby, the expected binomial moments of mRNA counts subject to extrinsic noise can be formulated as
\begin{equation}\label{noiseB}
    \langle B_m\rangle_\rho=\int_\Omega B_m(\bm{\theta})\rho(\bm{\theta}) \mathrm{d}\bm{\theta},
\end{equation}
where $\bm{\theta}$ abbreviates the vector of $2N^2-N+1$ parameters in \eqref{Reaction} and $\Omega$ denotes the parameter space $[0,\infty]^{2N^2-N+1}$. Substituting \eqref{noiseB} into \eqref{distribution}, the perturbed probability mass function can be reconstructed. For specific $\rho(\bm{\theta})$ and some simplified models, \eqref{noiseB} may be reduced to closed forms. In general, \eqref{noiseB} can be numerically evaluated by Monte Carlo integration since the computation of $B_m$ is relatively fast. 

This framework can be used to analyze the joint effects of intrinsic and extrinsic noise, which has been performed for some special cases of \eqref{Reaction} \cite{ExtrinsicNoiseHeavyTailed2020Phys.Rev.Lett.,Exactlysolvablemodels2020J.Chem.Phys.}. 

One main statement of the Heavy-Tailed Law is intrinsic noise alone cannot generate heavy-tailed mRNA distributions, meaning that the moment generating function $\mathcal{M}\left[\{\mathcal{P}_n\}_{n\in\mathbb{N}}\right](t):=\sum_{n=0}^\infty \mathcal{P}_n\mathrm{e}^{nt}$ diverges for any $t>0$. The Heavy-Tailed Law is originally proposed for the Leaky Telegraph model \cite{ExtrinsicNoiseHeavyTailed2020Phys.Rev.Lett.} and a generalization equivalent to the special case in \eqref{Reaction} where $D_1$ is diagonal \cite{Exactlysolvablemodels2020J.Chem.Phys.}. However, a mathematical proof is given only for the Leaky Telegraph model \cite{ExtrinsicNoiseHeavyTailed2020Phys.Rev.Lett.} and the $2^m$-multistate models proposed in \cite{Exactlysolvablemodels2020J.Chem.Phys.}. Here we provide a novel and elegant proof for the heavy tailed statement in \eqref{Reaction}. Denote the Poisson distribution with parameter $\Lambda$ by $\text{Pois}(\Lambda)$. We prove the following inequality using \eqref{Bound1}.  
\begin{align}\label{neq}
    \mathcal{M}\left[\{\mathcal{P}_n\}_{n\in\mathbb{N}}\right](t)\leq \mathrm{e}^{2\lVert D_1\rVert_{\infty}/\delta}\times\mathcal{M}\left[\text{Pois}(\frac{\lVert D_1\rVert_\infty}{\delta})\right](t). 
\end{align}
For Poisson distribution, the moment generating function is finite for any $t>0$, therefore so is $\{\mathcal{P}_n\}_{n\in\mathbb{N}}$ according to \eqref{neq}. It follows that a experimentally observed heavy-tailed distribution of mRNA copy number implies combined effects of intrinsic and extrinsic noise \cite{ExtrinsicNoiseHeavyTailed2020Phys.Rev.Lett.}. The proof of \eqref{neq} is in the Appendix, Section J. We conjecture \eqref{neq} can be sharpened to $\mathcal{M}\left[\{\mathcal{P}_n\}_{n\in\mathbb{N}}\right](t)\leq \mathcal{M}\left[\text{Pois}(\lVert D_1\rVert_\infty/\delta)\right](t)$, but this remains unproven.

\section{Discussion}
Typically, the analysis of stochastic transcription models serves three primary biophysical objectives: revealing transcriptional mechanisms by statistically fitting the model to experimental data; theoretically understanding the origin and regulation of noise at transcription level; and, ultimately, identifying general principles governing the dynamics of biophysical processes.

By fitting the theoretically derived probability distribution of mRNA counts to real-world experimental data, one may estimate the parameters in \eqref{Reaction}. However, statistical inference using single-cell snapshot data is another long-standing challenging problem in this area \cite{Inferringkineticsstochastic2013GenomeBiol.,Inferringtranscriptionalbursting2023R.Soc.OpenSci.}, and we do not attempt to discuss it comprehensively here. In principle, the model and results in this article may help boost relevant research. For example, only models with rather specific structures were considered previously due to the lack of theoretical support, but our work may enable statistical studies to be carried out within a general setting, allowing much flexibility in the model selection. We note that, to validate the applicability of our results in statistical inference, further investigation is needed to carefully design the inference framework, incorporate physical evidence to control the number of independent parameters in \eqref{Reaction}, and perform experiments using synthetic or experimental data.

To theoretically understand the impact of intrinsic and extrinsic noise in transcription, deeper analysis of results presented in the last subsection of RESULTS is required. For instance, the dependence of Fano factor \eqref{fano} and intrinsic noise \eqref{cv} on different parameters in \eqref{Reaction} should be investigated. How extrinsic perturbation twists the original probability distribution should also be studied based on \eqref{noiseB}. 

General principles governing gene expression remain the ultimate objective of biophysical research. The Heavy-Tailed Law serves as one such concise principle. In this work, we extend one main statement of the Heavy-Tailed Law to our model and provide an elegant proof, using techniques from functional analysis. The Heavy-Tailed Law, to some extent, clarifies the boundary of the CME framework. Within this framework, the steady-state distribution of mRNA copy number cannot be overly pathological, always decaying to zero faster than a Poisson distribution. On the one hand, observing non-heavy-tailed distributions in experimental data indicates the limitation of CME-based models, and calls for incorporating extrinsic noise, or even developing an entirely new framework of stochastic modeling. On the other hand, both numerical results and theoretical analysis based on CMEs should be governed the Heavy-Tailed Law, enabling a quick validation. In addition, the proof presented here may be of particular interest to theorists, as some abstract techniques from functional analysis are introduced and the proof is surprisingly elegant.

Note also that our model is theoretically motivated. When modeling a specific type of gene, physical background should be exploited to constrain the model complexity appropriately. A detailed exploration of gene-state design is left for future investigation.

In the end, we discuss the generalizability of the method in this article. Note that only transcription is considered in our model, while other critical biological processes including translation \cite{Exactdistributionsstochastic2022Phys.Rev.E,Analyticaldistributionsstochastic2008Proc.Natl.Acad.Sci.U.S.A.} and feedback loops \cite{Stochasticmodelingautoregulatory2020Biophys.J.a,Smallproteinnumber2020J.Chem.Phys.} are not incorporated. In principle, generalization to these more complicated models will typically fail. Based on our analysis, the conceptually constructive step is to find the solution to a partial differential equation system \eqref{CME2}, and this step is ad hoc. However, interestingly, we have recently found that generalization to complete stochastic gene expression models considering also translation may be possible with some extra techniques. We will report relevant results in our next work.

\section{Conclusion}
Stochastic gene expression is one of the central problems in biophysics, and numerous mathematical methods have been developed to provide insights into specific models. This article establishes a broad and unified framework for quantitatively investigating the stochastic transcription kinetics and related problems. While pursuing the utmost theoretical generality, we derive the exact time-dependent solution to the corresponding CME, thus demonstrating the solvability of CMEs in transcription models. While the analytical solution is, in general, not suitable for numerical computation, it lays the foundation for further theoretical developments and invites the use of more advanced techniques from analysis. For example, by calculating the temporal limit of the time-dependent dynamics, we recover the well-known matrix-form expression for steady-state binomial moments. Two novel inequalities for binomial moments and the probability mass function are derived using techniques from functional analysis. 
Specifically, we find that the probability distribution of mRNA copy number is upper-bounded by a constant multiple of a Poisson distribution, and that binomial moments are bounded from above in terms of the corresponding-order Taylor coefficients of the exponential function expanded around zero and evaluated at $\lVert D_1\rVert_{\infty}/\delta$. These two inequalities serve to (i) extend and elegantly prove the main conclusion of the Heavy-Tailed Law, (ii) estimate the truncation error in numerical computation, (iii) guide the design of general-purpose numerical methods including FSP, and (iv) rigorously establish the convergence of \eqref{distribution}. Notably, mathematical techniques developed here are of independent interest and applicable to other stochastic modeling problems, especially those used in the proofs of \eqref{converge} and \eqref{Bound1}. We numerically verify the analytical results using SSA and FSP, and then demonstrate the consistency of our work with two special cases, namely, the Telegraph model and the Markovian model under renewal condition. 
We also briefly analyze the combined effects of intrinsic and extrinsic noise. 
Compared with several widely-used methods, our result features both generality and low computational complexity.

\section{Data and Code Availability}
The Python code implementing numerical methods and generating figures during the study is available at \url{https://github.com/yuntao2022/Stochastic_Transcription}.

% \section{Appendix}
% Appendix can be found online.

\section{Author Contributions}
Y.L. and Y.Z. designed research; Y.L. performed research; and Y.L. and Y.Z. wrote the paper.

\section{Acknowledgments}
Y.L. was supported by Natural Science Foundation of China (125B10002), FDUROP (Fudan Undergraduate Research Opportunities Program) (24260), and Shanghai Undergraduate Training Program on Innovation and Entrepreneurship grant (S202510246532). Y.Z. was supported by National Key R$\&$D Program of China (2024YFA1012401), the Science and Technology Commission of Shanghai Municipality (23JC1400501), and Natural Science Foundation of China (12241103).

% \section*{Declaration of Interests}
% The authors declare no competing interest.

\appendix
\renewcommand{\theequation}{S.\arabic{equation}}
\setcounter{equation}{0}
\newpage
\section{Chemical Master Equation \eqref{CME2} and its Solution \eqref{generatingFunction}}\label{app1}
Since $a_{i,i}=-\sum_{\substack{k=1\\k\neq i}}^Na_{i,k}-\sum_{k=1}^Nb_{i,k}$, we rearrange the terms on the right of \eqref{CME} and formulate the CME as
\begin{equation}\label{CME3}
\begin{aligned}
    \frac{\partial}{\partial t}\mathbb{P}_{i,j}(m;t)=&\sum_{k=1}^Na_{k,j}\mathbb{P}_{i,k}(m;t)+\sum_{k=1}^Nb_{k,j}\mathbb{P}_{i,k}(m-1;t)\\&+(m+1)\delta\mathbb{P}_{i,j}(m+1;t)-m\delta\mathbb{P}_{i,j}(m;t).
\end{aligned}
\end{equation}
\eqref{CME3} holds for any $1\leq i,j\leq N$ and $m\in\mathbb{N}$. Recall that $\mathbb{P}_{i,j}(-1;t)=0$. Arrange $\mathbb{P}_{i,j}(m;t)\;\;(1\leq i,j\leq N)$ as a $N\times N$ matrix $\mathbb{P}(m;t)$ whose $(i,j)$-th entry is $\mathbb{P}_{i,j}(m;t)$, then it follows that 
\begin{equation}
\begin{aligned}
    \frac{\partial}{\partial t}\mathbb{P}(m;t)=\mathbb{P}(m;t)D_0+\mathbb{P}(m-1;t)D_1+(m+1)\delta\mathbb{P}(m+1;t)-m\delta\mathbb{P}(m;t).
\end{aligned}
\end{equation}
Thereby the matrix-form generating function $\mathcal{G}(t,z)=\sum_{k=0}^\infty\mathbb{P}(k;t)z^k,\,\mid z\mid\leq1$ satisfies partial differential system \eqref{CME2}.

We now show \eqref{CME2} is exactly solvable by presenting its unique solution with initial condition $\mathcal{G}(0,z)=\bm{I}_N$, namely, 
\begin{equation}\label{generatingFunction}
\begin{aligned}
    \mathcal{G}(t,z)&=\mathrm{e}^{D_0t}+\int_{\Omega_1}\left[1+(z-1)\mathrm{e}^{-\delta(t-t_1)}\right]\mathrm{e}^{D_0t_1}D_1\mathrm{e}^{D_0(t-t_1)}\mathrm{d}t_1\\&+\int_{\Omega_2}\prod_{i=1}^2\left[1+(z-1)\mathrm{e}^{-\delta(t-t_i)}\right]\mathrm{e}^{D_0t_1}D_1\mathrm{e}^{D_0(t_2-t_1)}D_1\mathrm{e}^{D_0(t-t_2)}\mathrm{d}\bm{t}\\&+\sum_{k=3}^\infty\int_{\Omega_k}\prod_{i=1}^k\left[1+(z-1)\mathrm{e}^{-\delta(t-t_i)}\right]\mathrm{e}^{D_0t_1}D_1\mathrm{e}^{D_0(t_2-t_1)}D_1\cdots \mathrm{e}^{D_0(t_k-t_{k-1})}D_1\mathrm{e}^{D_0(t-t_k)}\mathrm{d}\bm{t}.
\end{aligned}
\end{equation}

Convergence of \eqref{generatingFunction} can be verified.

Given analytical expression \eqref{generatingFunction}, the verification is straightforward by direct differentiation. Obviously $\mathcal{G}(t,z)$ given by \eqref{generatingFunction} satisfies the initial condition $\mathcal{G}(0,z)=\bm{I}_N$. Therefore we only need to calculate $\frac{\partial}{\partial t}\mathcal{G}(t,z)$ and $\frac{\partial}{\partial z}\mathcal{G}(t,z)$, respectively, and show the equation \eqref{CME2} holds. The calculation is tedious and we proceed in three steps. 

\subsection{Calculation of $\frac{\partial}{\partial t}\mathcal{G}(t,z)$}
Calculating $\frac{\partial}{\partial t}\mathcal{G}(t,z)$ is non-trivial since the integral domain in \eqref{generatingFunction} is also a function of $t$. Recall that $\Omega_k:=\{(t_1,t_2,\cdots,t_k)\mid0\leq t_1\leq t_2\leq\cdots\leq t_k\leq t\}\subseteq\mathbb{R}^k$.
For convenience, let $\mathcal{F}(t,k,\bm{t},z)$ denote the integrand in \eqref{generatingFunction} (Bold letter $\bm{t}$ abbreviates the variable of integration in \eqref{generatingFunction}, namely, the tuple $(t_1,t_2,\cdots,t_k)$).
\begin{equation}
\begin{aligned}
&\mathcal{F}(t,k,\bm{t},z):=
    \prod_{i=1}^k\left[1+(z-1)\mathrm{e}^{-\delta(t-t_i)}\right]\mathrm{e}^{D_0t_1}D_1\mathrm{e}^{D_0(t_2-t_1)}D_1\cdots \mathrm{e}^{D_0(t_k-t_{k-1})}D_1\mathrm{e}^{D_0(t-t_k)}.
\end{aligned}
\end{equation}
Then
\begin{equation}
\begin{aligned}
    \frac{\partial}{\partial t}\mathcal{G}(t,z)&=\mathrm{e}^{D_0t}D_0+\sum_{k=1}^\infty\frac{\partial}{\partial t}\int_{\Omega_k}\mathcal{F}(t,k,\bm{t},z)\mathrm{d}\bm{t}.
\end{aligned}
\end{equation}
The integral domain $\Omega_k$ varies with time $t$, hence, we cannot directly interchange the integral and the time derivative. Instead, we must apply the differentiation rule for integrals with variable parameters. 
\begin{equation}
\begin{aligned}
    \frac{\partial}{\partial t}\int_{\Omega_k}\mathcal{F}(t,k,\bm{t},z)\mathrm{d}\bm{t}&=\frac{\partial}{\partial t}\left[\int_0^t\int_{t_1}^t\cdots\int_{t_{k-1}}^t\mathcal{F}(t,k,\bm{t},z)\mathrm{d}t_k\cdots \mathrm{d}t_2\mathrm{d}t_1\right]\\
    &=\int_0^t\frac{\partial}{\partial t}\left[\int_{t_1}^t\cdots\int_{t_{k-1}}^t\mathcal{F}(t,k,\bm{t},z)\mathrm{d}t_k\cdots \mathrm{d}t_2\right]\mathrm{d}t_1\\&\quad\quad\quad+\left[\int_{t_1}^t\dots\int_{t_{k-1}}^t\mathcal{F}(t,k,\bm{t},z)\mathrm{d}t_k\cdots \mathrm{d}t_2\right]\bigg|_{t_1=t}\\
    &=\int_0^t\frac{\partial}{\partial t}\left[\int_{t_1}^t\cdots\int_{t_{k-1}}^t\mathcal{F}(t,k,\bm{t},z)\mathrm{d}t_k\cdots \mathrm{d}t_2\right]\mathrm{d}t_1\\
    &=\cdots\\
    &=\int_0^t\int_{t_1}^t\cdots\frac{\partial}{\partial t}\left[\int_{t_{k-1}}^t\mathcal{F}(t,k,\bm{t},z)dt_k\right]\mathrm{d}t_{k-1}\cdots \mathrm{d}t_1\\
    &=\int_{\Omega_k}\frac{\partial}{\partial t}\mathcal{F}(t,k,\bm{t},z)\mathrm{d}\bm{t}+\int_{\Omega_{k-1}}\mathcal{F}(t,k,\bm{t},z)\bigg|_{t_k=t}\mathrm{d}\bm{t}.
\end{aligned}
\end{equation}
Consider the first term on the right. We have  
\begin{equation}\label{prop1}
\begin{aligned}
    \frac{\partial}{\partial t}\mathcal{F}(t,k,\bm{t},z)&=\sum_{i=1}^k(1-z)\delta\mathrm{e}^{-\delta(t-t_i)}\prod_{\substack{j=1\\j\neq i}}^k\left[1+(z-1)\mathrm{e}^{-\delta(t-t_j)}\right]\\&\mathrm{e}^{D_0t_1}D_1\mathrm{e}^{D_0(t_2-t_1)}D_1\cdots\mathrm{e}^{D_0(t_k-t_{k-1})}D_1\mathrm{e}^{D_0(t-t_k)}\\&+\prod_{i=1}^k\left[1+(z-1)\mathrm{e}^{-\delta(t-t_i)}\right]\\&\mathrm{e}^{D_0t_1}D_1\mathrm{e}^{D_0(t_2-t_1)}D_1\cdots \mathrm{e}^{D_0(t_k-t_{k-1})}D_1\mathrm{e}^{D_0(t-t_k)}D_0.
\end{aligned}
\end{equation}
On the other hand, 
\begin{equation}\label{prop2}
\begin{aligned}
    \int_{\Omega_{k-1}}\mathcal{F}(t,k,\bm{t},z)\bigg|_{t_k=t}\mathrm{d}\bm{t}&=\int_{\Omega_{k-1}}z\prod_{i=1}^{k-1}\left[1+(z-1)\mathrm{e}^{-\delta(t-t_i)}\right]\\&\quad\quad\quad\quad\mathrm{e}^{D_0t_1}D_1\mathrm{e}^{D_0(t_2-t_1)}D_1\cdots\mathrm{e}^{D_0(t-t_{k-1})}D_1\mathrm{d}\bm{t}.
\end{aligned}
\end{equation}
In particular, when $k=1$, the right hand side is taken as $z\mathrm{e}^{D_0t}D_1$. 

\subsection{Calculation of $\frac{\partial}{\partial Z}\mathcal{G}(t,z)$}
Calculating $\frac{\partial}{\partial Z}\mathcal{G}(t,z)$ is standard. Observe that
\begin{equation}\label{prop3}
\begin{aligned}
    \frac{\partial}{\partial z}\mathcal{G}(t,z)&=\sum_{k=1}^\infty\int_{\Omega_k}\frac{\partial}{\partial z}\prod_{i=1}^k\left[1+(z-1)\mathrm{e}^{-\delta(t-t_i)}\right]\\&\quad\quad\quad\quad\quad\quad\quad\mathrm{e}^{D_0t_1}D_1\mathrm{e}^{D_0(t_2-t_1)}D_1\cdots \mathrm{e}^{D_0(t_k-t_{k-1})}D_1\mathrm{e}^{D_0(t-t_k)}\mathrm{d}\bm{t}\\
    &=\sum_{k=1}^\infty\int_{\Omega_k}\sum_{j=1}^k\mathrm{e}^{-\delta(t-t_j)}\prod_{\substack{i=1\\i\neq j}}^k\left[1+(z-1)\mathrm{e}^{-\delta(t-t_i)}\right]\\&\quad\quad\quad\quad\quad\quad\quad \mathrm{e}^{D_0t_1}D_1\mathrm{e}^{D_0(t_2-t_1)}D_1\cdots\mathrm{e}^{D_0(t_k-t_{k-1})}D_1\mathrm{e}^{D_0(t-t_k)}\mathrm{d}\bm{t}.
\end{aligned}
\end{equation}
Substituting \eqref{prop1}, \eqref{prop2} into the left hand side of \eqref{CME2} and \eqref{prop3} into the right hand side of \eqref{CME2}, it is easily seen that equation \eqref{CME2} holds. 

\subsection{Analytical Expression of $\mathbb{P}_{i,j}(m;t)$}
To obtain exact time-dependent solution to the CME \eqref{CME}, namely, the analytical expression of $\mathbb{P}_{i,j}(m;t)$, we arrange the series \eqref{generatingFunction} according to powers of $z$.
For $m\geq 1$, we have
\begin{equation}\label{density3}
\begin{aligned}
\mathbb P(m;t)
&= \sum_{k=m}^{\infty} \int_{\Omega_k}
   \sum_{\substack{I\subseteq\{1,\cdots,k\}\\ \# I=m}}
   \Bigg[\prod_{i\in I}\alpha_i \prod_{j\in I^{\mathrm c}} (1-\alpha_j) \Bigg]
   \mathrm e^{D_0 t_1} D_1 \mathrm e^{D_0(t_2-t_1)} D_1 \cdots
   D_1 \mathrm e^{D_0(t-t_k)}\mathrm d\bm{t};
\end{aligned}
\end{equation}
when $m=0$, we have
\begin{equation}\label{density4}
\begin{aligned}
\mathbb{P}(0;t)=&\mathrm{e}^{D_0t}+\sum_{k=1}^\infty\int_{\Omega_k}\Bigg[\prod_{j=1}^{k}\bigl(1-\alpha_j\bigr)\Bigg]
    \mathrm{e}^{D_0t_1}D_1\mathrm{e}^{D_0(t_2-t_1)}D_1\cdots D_1\mathrm{e}^{D_0(t-t_k)}\mathrm{d}\bm{t}.
\end{aligned}
\end{equation}
Notations in \eqref{density3} and \eqref{density4} are explained in the main text, right after \eqref{density} and \eqref{density0}.
Recall that $\mathbb{P}_{i,j}(m;t)$ is the $(i,j)$-th entry of matrix $\mathbb{P}(m;t)$, therefore \eqref{density} and \eqref{density0} follow.

\section{Derivation of Limiting Binomial Moments}\label{app2}
\subsection{Time-dependent Binomial Moments}
We note that the proof in this subsection essentially follows \cite{AnalysisInfiniteServerQueue2002QueueingSyst.}.

The matrix-form binomial moment $\mathcal{B}(m,t)$ is defined (entry-wise) as the $m$-th coefficient in the Taylor expansion of $\mathcal{G}(t,z)$ at $z=1$ (should it exists). Let $w=z-1$.
To obtain the analytical expression of $\mathcal{B}(m,t)$, we first prove the following equality
\begin{align}
\mathcal{G}(t,z)&\equiv\mathrm{e}^{D_0t}+\sum_{k=1}^\infty\int_{\Omega_k}\prod_{i=1}^k\left[1+w\mathrm{e}^{-\delta(t-t_i)}\right]\mathrm{e}^{D_0t_1}D_1\mathrm{e}^{D_0(t_2-t_1)}D_1\cdots \mathrm{e}^{D_0(t_k-t_{k-1})}D_1\mathrm{e}^{D_0(t-t_k)}\mathrm{d}\bm{t}\label{G1}
\\&=
\mathrm{e}^{Dt}+\sum_{k=1}^\infty\int_{\Omega_k}w^k\mathrm{e}^{-\delta(kt-\sum_{j=1}^kt_j)}\mathrm{e}^{Dt_1}D_1\mathrm{e}^{D(t_2-t_1)}\cdots\mathrm{e}^{D(t_{k}-t_{k-1})}D_1\mathrm{e}^{D(t-t_k)}\mathrm{d}\bm{t}.\label{G2}
\end{align}
We prove \eqref{G1} and \eqref{G2} are the same by using uniqueness of solution to a partial differential equation system.
According to \eqref{CME2} and \eqref{generatingFunction}, \eqref{G1} satisfies
\begin{equation}\label{dynamics2}
\begin{aligned}
\frac{\partial}{\partial t}\mathcal{G}(t,w)=\mathcal{G}(t,w)D+w\mathcal{G}(t,w)D_1-w\delta\frac{\partial}{\partial w}\mathcal{G}(t,w).
\end{aligned}
\end{equation}
We then verify \eqref{G2} also satisfies \eqref{dynamics2}, with initial condition $\mathcal{G}(0,w)=\bm{I}_N$. 
The proof is an analogue of the one in \autoref{app1} and we omit it here. Therefore, we prove the equivalence of \eqref{G1} and \eqref{G2}. 
Perform change of variables $x_i=t-t_{k+1-i}(i=1,2,\cdots,k)$, \eqref{G2} can also be written as
\begin{equation}\label{generatingbinomial2}
\begin{aligned}
\mathcal{G}(t,w)&=\mathrm{e}^{Dt}+\sum_{k=1}^\infty w^k\int_{\Omega_k}\mathrm{e}^{-\delta\sum_{j=1}^kx_j}\mathrm{e}^{D(t-x_k)}D_1\mathrm{e}^{D(x_k-x_{k-1})}\cdots\mathrm{e}^{D(x_2-x_1)}D_1\mathrm{e}^{Dx_1}\mathrm{d}\bm{x}.
\end{aligned}
\end{equation}
Since \eqref{generatingbinomial2} naturally takes the form of a power series, we obtain an expression of binomial moments ($m\geq1$)
\begin{equation}\label{BinomialRaw}
\begin{aligned}
    \mathcal{B}(m,t)=\int_{\Omega_m}\mathrm{e}^{-\delta\sum_{j=1}^mx_j}\mathrm{e}^{D(t-x_m)}D_1\mathrm{e}^{D(x_m-x_{m-1})}\cdots\mathrm{e}^{D(x_2-x_1)}D_1\mathrm{e}^{Dx_1}\mathrm{d}\bm{x}.
\end{aligned}
\end{equation}
In particular, $\mathcal{B}(0,t)=\mathrm{e}^{Dt}$.
\subsection{Take Temporal Limit}
Recall that by summing over current states and setting the initial distribution to be $\bm{\pi}$, we define scalar binomial moments $B_m(t):=\bm{\pi}^\top\mathcal{B}(m,t)\bm{1}$ and $\bm{\pi}$ is uniquely determined by $\bm{\pi}^\top D=\bm{0}_{1\times N}$ and $D\bm{1}=\bm{0}_{N\times 1}$.
In this subsection, we prove 
\begin{equation}\label{BinomialTarget}
\begin{aligned}
    &\lim_{t\rightarrow\infty}\int_{\Omega_m}\mathrm{e}^{-\delta\sum_{j=1}^mx_j}\bm{\pi}^\top D_1\mathrm{e}^{D(x_m-x_{m-1})}\cdots \mathrm{e}^{D(x_2-x_1)}D_1\bm{1}\mathrm{d}\bm{x}\\=&
    \frac{1}{m\delta}\bm{\pi}^\top D_1\left(\delta\bm{I}_N-D\right)^{-1}D_1\left(2\delta\bm{I}_N-D\right)^{-1}\cdots D_1
   \left[(m-1)\delta\bm{I}_N-D\right]^{-1}D_1\bm{1},\; m\geq 2;
\end{aligned}
\end{equation}
and
\begin{equation}\label{BinomialTarget1}
\begin{aligned}
 \lim_{t\rightarrow\infty}\int_0^t\mathrm{e}^{-\delta x_1}\bm{\pi}^\top D_1\bm{1}\mathrm{d}x_1
    =\frac{1}{\delta}\bm{\pi}^\top D_1\bm{1}.
\end{aligned}
\end{equation}

Since \eqref{BinomialTarget1} is obvious, we only consider \eqref{BinomialTarget} now.

By L\'{e}vy-Desplanques theorem, any strictly diagonally dominant matrix is nonsingular, which in particular applies to $k\delta\bm{I}_N-D$ when $\delta>0,k\geq1$ and $D$ is a $Q$-matrix. By Ger\v{s}gorin disc theorem, we further conclude that any eigenvalue of a strictly diagonally dominant matrix with negative diagonal entries has negative real part. 

Let $B_m:=\lim_{t\rightarrow\infty}B_m(t)$ be the stationary $m$-th binomial moment of mRNA copy number in the model \eqref{Reaction} (should it converges). We now prove \eqref{binomial}.
According to \eqref{BinomialTarget}, 
\begin{equation}\label{binomiallimit}
\begin{aligned}
    B_m&=\lim_{t\rightarrow\infty}B_m(t)\\
    &=\int_0^\infty\int_0^{x_m}\cdots\int_0^{x_2}\mathrm{e}^{-\delta\sum_{j=1}^mx_j}\bm{\pi}^\top D_1\mathrm{e}^{D(x_m-x_{m-1})}\cdots\mathrm{e}^{D(x_2-x_1)}D_1\bm{1}\mathrm{d}x_1\mathrm{d}x_2\cdots \mathrm{d}x_{m-1}\mathrm{d}x_m,
\end{aligned} 
\end{equation}
thus $B_m$ exists as long as the integral above converges. 
By Fubini theorem,
\begin{equation}
\begin{aligned}
    B_m=&\int_0^\infty\int_{x_1}^\infty\cdots\int_{x_{m-1}}^\infty\mathrm{e}^{-\delta\sum_{j=1}^mx_j}\bm{\pi}^\top D_1\mathrm{e}^{D(x_m-x_{m-1})}\cdots\mathrm{e}^{D(x_2-x_1)}D_1\bm{1}\mathrm{d}x_m\cdots \mathrm{d}x_2\mathrm{d}x_1.
\end{aligned}
\end{equation}
Note that, for $k\geq1$ and $\delta>0$, since $k\delta\bm{I}_N-D$ is nonsingular as we have pointed out, 
\begin{equation}\label{integral}
\begin{aligned}
 &\int_{x_{i-1}}^\infty\mathrm{e}^{-k\delta x_i}\mathrm{e}^{D(x_i-x_{i-1})} \mathrm{d}x_i\\
 =&\int_{x_{i-1}}^\infty\mathrm{e}^{(D-k\delta\bm{I}_N)x_i-Dx_{i-1}} \mathrm{d}x_i\\
 =&\left(D-k\delta\bm{I}_N\right)^{-1}\mathrm{e}^{(D-k\delta\bm{I}_N)x_i}\bigg|_{x_i=x_{i-1}}^\infty\mathrm{e}^{-Dx_{i-1}}\\
 =&\left(k\delta\bm{I}_N-D\right)^{-1}\mathrm{e}^{(D-k\delta\bm{I}_N)x_{i-1}}\mathrm{e}^{-Dx_{i-1}}\\
 =&\left(k\delta\bm{I}_N-D\right)^{-1}\mathrm{e}^{-k\delta x_{i-1}}.
\end{aligned}
\end{equation}
In the penultimate equality, we have used $\lim_{t\rightarrow\infty} \mathrm{e}^{(D-k\delta\bm{I}_N)t}=\bm{0}_{N\times N}$. This can be shown by transforming $D-k\delta\bm{I}_N$ into the Jordan canonical form and observe that eigenvalues of $D-k\delta\bm{I}_N$ all have strictly negative real part (see Chapter $9$ in \cite{MatrixComputations2013}). Repeatedly employing \eqref{integral}, we obtain eventually \eqref{binomial}. Besides, since the right hand side of \eqref{binomiallimit} is well defined, $\{B_m\}_{m\in\mathbb{N}}$ exists.

\section{Upper Bound for Binomial Moments \eqref{converge}}\label{app3}
To further analyze the asymptotic behavior of $\{B_m\}_{m\in\mathbb{N}}$, we exploit an analogue of Theorem 4.1.2 in \cite{MatrixComputations2013}, which states that: 
\begin{theorem}\label{fromMC}
Let $C=(c_{i,j})_{N\times N}$ be a $N\times N$ strictly row diagonally dominant matrix with $\Theta:=\min_{1\leq i\leq N}\left(\mid c_{i,i}\mid-\sum_{j\neq i}\mid c_{i,j}\mid\right)$. Then $\lVert C^{-1}\rVert_\infty\leq \Theta^{-1}$. 
\end{theorem}
This theorem provides an upper bound with respect to infinite norm for the inverse of one diagonally dominant matrix. 
According to the above theorem, $\lVert \left(k\delta\bm{I}_N-D\right)^{-1}\rVert_\infty\leq (k\delta)^{-1}$ for $k\geq1$. By definition of operator norm $\lVert Ax\rVert_\infty\leq\lVert A\rVert_\infty\lVert x\rVert_\infty$ and $\lVert AB\rVert_\infty\leq\lVert A\rVert_\infty\lVert B\rVert_\infty$. Therefore, we get ($m\geq1$)
\begin{equation}
\begin{aligned}
    B_m&= \frac{1}{m\delta}\bm{\pi}^\top D_1\left(\delta\bm{I}_N-D\right)^{-1}D_1\left(2\delta\bm{I}_N-D\right)^{-1}\cdots D_1\left[(m-1)\delta\bm{I}_N-D\right]^{-1}D_1\bm{1}\\
    &\leq\frac{1}{m\delta}\lVert\bm{\pi}\rVert_1\lVert D_1\left(\delta\bm{I}_N-D\right)^{-1}D_1\left(2\delta\bm{I}_N-D\right)^{-1}\cdots D_1\left[(m-1)\delta\bm{I}_N-D\right]^{-1}D_1\bm{1}\rVert_\infty\\
    &\leq\frac{1}{m\delta}\lVert\bm{\pi}\rVert_1\lVert D_1\rVert_\infty\lVert \left(\delta\bm{I}_N-D\right)^{-1}\rVert_\infty\cdots\lVert \left[(m-1)\delta\bm{I}_N-D\right]^{-1}\rVert_\infty\lVert D_1\rVert_\infty\lVert \bm{1}\rVert_\infty\\
    &\leq\frac{1}{m\delta}\lVert\bm{\pi}\rVert_1\lVert D_1\rVert_\infty\frac{1}{\delta}\lVert D_1\rVert_\infty\cdots\lVert D_1\rVert_\infty\frac{1}{(m-1)\delta}\lVert D_1\rVert_\infty\lVert \bm{1}\rVert_\infty\\
    &=\frac{1}{m!}\left(\frac{\lVert D_1\rVert_{\infty}}{\delta}\right)^m.
\end{aligned}
\end{equation}
In the first inequality, H\"older inequality is employed. 

Since for any given $D_1$ and $\delta>0$
\begin{equation}
    \lim_{m\rightarrow\infty}\frac{1}{m!}\left(\frac{\lVert D_1\rVert_{\infty}}{\delta}\right)^m=0, 
\end{equation}
we also conclude that $\{B_m\}_{m\in\mathbb{N}}$ tends to zero. 

\section{Upper Bound for Probability Mass Function \eqref{Bound1}}\label{app4}
Accordingly, we reconstruct $\{\mathcal{P}_n\}_{n\in\mathbb{N}}$ by applying \eqref{distribution}.
Identity \eqref{distribution} can be proved as follows.
Recall that 
\begin{equation}
\begin{aligned}
    \mathcal{P}_n:=\lim_{t\rightarrow\infty}\bm{\pi}^\top\mathbb{P}(m;t)\bm{1}=\frac{1}{n!}\lim_{t\rightarrow\infty}\left[\frac{\partial^n}{\partial z^n}\bm{\pi}^\top\mathcal{G}(t,z)\bm{1}\bigg|_{z=0}\right],
\end{aligned}
\end{equation}
and
\begin{equation}
\begin{aligned}
     B_m:=\lim_{t\rightarrow\infty}\bm{\pi}^\top\mathcal{B}(m,t)\bm{1}= \frac{1}{m!}\lim_{t\rightarrow\infty}\left[\frac{\partial^m}{\partial z^m}\bm{\pi}^\top\mathcal{G}(t,z)\bm{1}\bigg|_{z=1}\right].
\end{aligned}
\end{equation}
Therefore
\begin{equation}
    \lim_{t\rightarrow\infty}\bm{\pi}^\top\mathcal{G}(t,z)\bm{1}=\sum_{n=0}^\infty\mathcal{P}_nz^n=\sum_{m=0}^\infty B_m(z-1)^m.
\end{equation}
By comparing the coefficients of $z^n$, we obtain \eqref{distribution}. Similarly, we can prove \eqref{distribution_reverse}.

We now prove \eqref{Bound1}.
By repeatedly applying \eqref{binomial}, 
\begin{equation}
\begin{aligned}
    \mathcal{P}_n&=\sum^{\infty}_{m=n}\left(-1\right)^{m-n}\binom{m}{n}B_m\\&\leq\sum^{\infty}_{m=n}\binom{m}{n} B_m\\
    &\leq \sum^{\infty}_{m=n}\binom{m}{n}\frac{1}{m!}\left(\frac{\lVert D_1\rVert_{\infty}}{\delta}\right)^m\\
    &=\frac{1}{n!}\sum^{\infty}_{m=n}\frac{1}{(m-n)!}\left(\frac{\lVert D_1\rVert_{\infty}}{\delta}\right)^m\\
    &=\frac{1}{n!}\left(\frac{\lVert D_1\rVert_{\infty}}{\delta}\right)^n\sum^{\infty}_{m=n}\frac{1}{(m-n)!}\left(\frac{\lVert D_1\rVert_{\infty}}{\delta}\right)^{m-n}\\
    &=\frac{1}{n!}\left(\frac{\lVert D_1\rVert_{\infty}}{\delta}\right)^n\sum^{\infty}_{k=0}\frac{1}{k!}\left(\frac{\lVert D_1\rVert_{\infty}}{\delta}\right)^k\\
    &=\frac{1}{n!}\left(\frac{\lVert D_1\rVert_{\infty}}{\delta}\right)^n\mathrm{e}^{\lVert D_1\rVert_{\infty}/\delta},
\end{aligned}
\end{equation}
which is exactly what we claim in \eqref{Bound1}. Since the right hand side of \eqref{Bound1} tends to zero, we also obtain that the sequence $\{\mathcal{P}_n\}_{n\in\mathbb{N}}$ converges to zero.

\section{Upper Bound for Truncation Error}\label{app5}
We now estimate the truncation error. 
Define $\widehat{\mathcal{P}_n}$ as the computed value given the truncation point $M$. Assume that $M>\max\{2n+2,8\mathrm{e}\left(\lVert D_1\rVert_{\infty}/\delta\right)^2\}$, where $\mathrm{e}$ is Euler's number. We now prove \eqref{bound2}.
According to \eqref{binomial}, \eqref{distribution} and apply the inequality 
\begin{equation}\label{fact}
n!\geq \left(\frac{n}{\mathrm{e}}\right)^n\sqrt{2\pi n},
\end{equation}
we obtain
\begin{equation}
\begin{aligned}
    \mid \mathcal{P}_n-\widehat{\mathcal{P}_n}\mid&=\mid\sum^{\infty}_{m=M}\left(-1\right)^{m-n}\binom{m}{n}B_m\mid\\
    &\leq \sum^{\infty}_{m=M}\binom{m}{n}B_m\\
    &\leq \frac{1}{n!}\sum^{\infty}_{m=M}\frac{1}{(m-n)!}\left(\frac{\lVert D_1\rVert_{\infty}}{\delta}\right)^m\\
    &\leq \frac{1}{n!}\sum^{\infty}_{m=M}\frac{1}{\lfloor m/2 \rfloor!}\left(\frac{\lVert D_1\rVert_{\infty}}{\delta}\right)^m\\
    &\leq\frac{1}{n!}\sum^{\infty}_{m=M}\left(\frac{\lfloor m/2 \rfloor}{\mathrm{e}}\right)^{-\lfloor m/2 \rfloor}\frac{1}{\sqrt{2\pi\lfloor m/2 \rfloor}}\left(\frac{\lVert D_1\rVert_{\infty}}{\delta}\right)^m\\
    &\leq\frac{1}{n!}\sum^{\infty}_{m=M}\left(\frac{m-1}{\mathrm{2e}}\right)^{-\frac{m-1}{2}}\frac{1}{\sqrt{(m-1)\pi}}\left(\frac{\lVert D_1\rVert_{\infty}}{\delta}\right)^m\\
    &= \frac{1}{n!}\sum^{\infty}_{m=M}\frac{(\sqrt{2\mathrm{e}})^{m-1}}{\sqrt{\pi}(\sqrt{m-1})^m}\left(\frac{\lVert D_1\rVert_{\infty}}{\delta}\right)^m\\
    &=\frac{1}{\sqrt{2\pi\mathrm{e}}n!}\sum^{\infty}_{m=M}\left(\frac{\sqrt{2\mathrm{e}}\lVert D_1\rVert_{\infty}}{\sqrt{m-1}\delta}\right)^m\\
    & \leq \frac{1}{\sqrt{2\pi\mathrm{e}}n!}\sum^{\infty}_{m=M}\frac{1}{2^m}\\
    &=\frac{1}{\sqrt{2\pi\mathrm{e}}n!2^{M-1}}. 
\end{aligned}
\end{equation}

\section{Special Case I: The Telegraph Model}\label{app6} 
By definition \begin{equation}
    \bm{\pi}=\left[k_2/(k_1+k_2),k_1/(k_1+k_2)\right]^\top
\end{equation}
and 
\begin{equation}
    \left(k\delta\bm{I}_2-D\right)^{-1}=\frac{1}{k\delta(k\delta+k_1+k_2)}
    \begin{pmatrix}
    k_{2} +k\delta & k_{1}\\
    k_{2} & k_{1} +k\delta 
    \end{pmatrix}.
\end{equation}
Substituting into \eqref{binomial}, we have
\begin{equation*}
\begin{aligned}
    B_m&=\frac{1}{m!}\left(\frac{\lambda}{\delta}\right)^m\frac{(k_2/\delta)^{(m)}}{(k_1/\delta+k_2/\delta)^{(m)}}.
\end{aligned}
\end{equation*}
where $x^{(m)}$ is the Pochhammer symbol, denoting the $m$-th rising factorial of $x$. Namely, $x^{(m)}:=\Gamma(x+m)/\Gamma(x)$ and $\Gamma(x)$ is Gamma function.
Substituting \eqref{telegraph} into \eqref{distribution},
\begin{equation}\label{1F1}
\begin{aligned}
    \mathcal{P}_n&=\sum^{\infty}_{m=n}(-1)^{m-n}\binom{m}{n}B_m\\
    &=\frac{1}{n!}\sum^{\infty}_{m=n}\frac{(-1)^{m-n}}{(m-n)!}\left(\frac{\lambda}{\delta}\right)^m\frac{(k_2/\delta)^{(m)}}{(k_1/\delta+k_2/\delta)^{(m)}}\\
    &=\frac{1}{n!}\left(\frac{\lambda}{\delta}\right)^n\frac{(k_2/\delta)^{(n)}}{(k_1/\delta+k_2/\delta)^{(n)}}\sum^{\infty}_{k=0}\frac{(k_2/\delta+n)^{(k)}}{(k_1/\delta+k_2/\delta+n)^{(k)}}\left(-\frac{\lambda}{\delta}\right)^k.
\end{aligned}
\end{equation}
Denote by \;$\mathllap{_{1}}F_1(a,b;z):=\sum_{n=0}^\infty\frac{a^{(n)}}{b^{(n)}}\frac{z^n}{n!}$ the confluent hypergeometric function, \eqref{1F1} can be formulated as \eqref{1F1s}
\begin{equation*}
\begin{aligned}
    \mathcal{P}_n&=\frac{1}{n!}\left(\frac{\lambda}{\delta}\right)^n\frac{(k_2/\delta)^{(n)}}{(k_1/\delta+k_2/\delta)^{(n)}}\quad\mathllap{_{1}}F_1\left(\frac{k_2}{\delta}+n,\frac{k_1+k_2}{\delta}+n;-\frac{\lambda}{\delta}\right).
\end{aligned}
\end{equation*}

\section{Special Case II: Markovian Models under Renewal Condition}\label{app7}
The Laplace transform of $f(t)$ can be easily reduced by substituting \eqref{pdf} into the definition of Laplace transform. 
\begin{equation}
\begin{aligned}
    \mathcal{L}[f](s)&=\int_0^\infty f(t)e^{-st}\;\mathrm{d}t\\&=\int_0^\infty\bm{\kappa}^\top\mathrm{e}^{(D_0-s\bm{I}_N)t}D_1\bm{1}\;\mathrm{d}t\\
    &=\bm{\kappa}^\top\left(\int_0^\infty\mathrm{e}^{(D_0-s\bm{I}_N)t}\;\mathrm{d}t\right)D_1\bm{1}\\
    &=\bm{\kappa}^\top(s\bm{I}_N-D_0)^{-1}D_1\bm{1},\;\;s>0.
\end{aligned}
\end{equation}
In the last equality, we have used $\lim_{t\rightarrow\infty} \mathrm{e}^{(D_0-s\delta\bm{I}_N)t}=\bm{0}_{N\times N}$ when $s>0$. Note that eigenvalues of $D_0-s\delta\bm{I}_N$ have strictly negative real part according to Ger\v{s}gorin disc theorem. 
Subsequently, \eqref{GIqueue} is reduced by analyzing single term and $\alpha$, receptively. 
We first prove the following equality. 
\begin{equation}\label{claim3}
    \frac{\mathcal{L}[f](s)}{1-\mathcal{L}[f](s)}=\bm{\kappa}^\top(s\bm{I}_N-D)^{-1}D_1\bm{1}.
\end{equation}
For $s>0$, both $s\bm{I}_N-D_0$ and $s\bm{I}_N-D$ are nonsingular by L\'{e}vy-Desplanques theorem. We assert that $(s\bm{I}_N-D_0)^{-1}=(s\bm{I}_N-D)^{-1}-(s\bm{I}_N-D_0)^{-1}D_1(s\bm{I}_N-D)^{-1}$. This can be seen by multiplying $(s\bm{I}_N-D_0)^{-1}$ on the left and and $(s\bm{I}_N-D)^{-1}$ on the right of the equality $s\bm{I}_N-D=(s\bm{I}_N-D_0)-D_1$, respectively. Thus, 
\begin{equation}
\begin{aligned}
    \bm{\kappa}^\top(s\bm{I}_N-D_0)^{-1}D_1\bm{1}&=\bm{\kappa}^\top(s\bm{I}_N-D)^{-1}D_1\bm{1}-\bm{\kappa}^\top(s\bm{I}_N-D_0)^{-1}D_1(s\bm{I}_N-D)^{-1}D_1\bm{1}\\
    &=\bm{\kappa}^\top(s\bm{I}_N-D)^{-1}D_1\bm{1}-\bm{\kappa}^\top(s\bm{I}_N-D_0)^{-1}D_1\bm{1}\bm{\kappa}^\top(s\bm{I}_N-D)^{-1}D_1\bm{1}\\
    &=\left[1-\bm{\kappa}^\top(s\bm{I}_N-D_0)^{-1}D_1\bm{1}\right]\bm{\kappa}^\top(s\bm{I}_N-D)^{-1}D_1\bm{1}.
\end{aligned}
\end{equation}
At the penultimate equality, we employed the equality $D_1=D_1\bm{1}\bm{\kappa}^\top$. Rearrange the above equality and we obtain \eqref{claim3}. 

We then give a further reduction of the expected inter-arrival time $\alpha$. To guarantee the convergence of $\alpha$, $D_0$ needs to be nonsingular. We claim that $D_0$ is always nonsigular, and the proof is placed in the next section.
We now prove that
\begin{equation}\label{meantime}
\alpha=-\bm{\kappa}^\top D_0^{-1}\bm{1}.
\end{equation}
Note that 
\begin{equation}
    \frac{\mathrm{d}}{\mathrm{d}t}\left(t\mathrm{e}^{D_0t}\right)=\mathrm{e}^{D_0t}+tD_0\mathrm{e}^{D_0t}.
\end{equation}
Thereby we have
\begin{equation}
    \begin{aligned}
        \int_0^\infty te^{D_0t}\;\mathrm{d}t&= D_0^{-1}\int_0^\infty \left[\frac{\mathrm{d}}{\mathrm{d}t}\left(t\mathrm{e}^{D_0t}\right)-\mathrm{e}^{D_0t}\right]\;\mathrm{d}t\\
        &=D_0^{-1}\left(t\mathrm{e}^{D_0t}\bigg|_{t=0}^\infty+D_0^{-1}\right)\\
        &=D_0^{-2}
    \end{aligned}
\end{equation}
Since $D_0$ is nonsingular and row diagonally dominant with off-diagonal entries nonnegative, diagonal entries negative, the real part of any eigenvalue of $D_0$ is strictly negative. In the last equality we employ $\lim_{t\rightarrow\infty}t\mathrm{e}^{D_0t}=\bm{0}_{N\times N}$ and $\lim_{t\rightarrow\infty}\mathrm{e}^{D_0t}=\bm{0}_{N\times N}$ (see Chapter $9$ in \cite{MatrixComputations2013}). 
Then 
\begin{equation}
\begin{aligned}
    \alpha&=\int_0^\infty \bm{\kappa}^\top t\mathrm{e}^{D_0t}D_1\bm{1}\;\mathrm{d}t\\
    &=\bm{\kappa}^\top\int_0^\infty t\mathrm{e}^{D_0t}\;\mathrm{d}t\;(-D_0)\bm{1}\\
    &=-\bm{\kappa}^\top D_0^{-1}\bm{1}.
\end{aligned}
\end{equation}
At the penultimate equality, we have applied $D\bm{1}=\bm{0}_{N\times 1}$.

Moreover, 
\begin{equation}\label{claim2}
    \frac{1}{\alpha}\bm{\kappa}^\top=\bm{\pi}^\top D_1.
\end{equation}
This is proved as follows. 
By \eqref{meantime} and the fact that $\bm{\pi}^\top D=\bm{0}_{1\times N}$, we get \begin{equation}
    \alpha\bm{\pi}^\top D_1=\bm{\kappa}^\top D_0^{-1}\bm{1}\bm{\pi}^\top D_0.
\end{equation}
We claim that 
\begin{equation}\label{claim}
    \bm{\kappa}^\top(\bm{I}_N-D_0^{-1}\bm{1}\bm{\pi}^\top D_0)=\bm{0}_{1\times N}.
\end{equation}
Abbreviate $\sum_{j=1}^N b_{i,j}$ as $\beta_i$ and $\bm{\beta}:=(\beta_1,\beta_2,\cdots,\beta_N)^\top$. Then 
\begin{equation}\label{D1}
     D_1=\bm{\beta}\bm{\kappa}^\top.
\end{equation}
Insert \eqref{D1} into $\bm{\pi}^\top(D_0+D_1)=0$ and multiply $D_0^{-1}$ on the right. We obtain 
\begin{equation}
\bm{\pi}^\top+\bm{\pi}^\top\bm{\beta}\bm{\kappa}^\top D_0^{-1}=\bm{0}_{1\times N}.
\end{equation}
Since $\bm{\pi}^\top\bm{\beta}$ is a positive scalar, 
\begin{equation}
     \bm{\kappa}^\top D_0^{-1}=-\frac{1}{\bm{\pi}^\top\bm{\beta}}\bm{\pi}^\top.
\end{equation}
Thereby, 
\begin{equation}
     \bm{\kappa}^\top D_0^{-1}\left(\bm{I}_N-\bm{1}\bm{\pi}^\top\right)=-\frac{1}{\bm{\pi}^\top\bm{\beta}}(\bm{\pi}^\top-\bm{\pi}^\top\bm{1}\bm{\pi}^\top)=\bm{0}_{1\times N}.
\end{equation}
Multiply $D_0$ on the right and we prove \eqref{claim} as claimed. \eqref{claim2} follows immediately. 

Substituting \eqref{claim3} and \eqref{claim2} into \eqref{GIqueue}, we arrive at \eqref{binomial}, proving the consistency of \eqref{GIqueue} and our result.

\section{Nonsingularity of $D_0$ under Renewal Condition}\label{appplus}
With the assumption $D$ is irreducible and $\text{rank}\,D_1=1$, we now prove $D_0$ is nonsingular. 

Without loss of generality, assume $N\geq 2$.

Denote
\begin{equation}
    D=:(d_{i,j})_{N\times N}=\begin{pmatrix} \bm{d}_1^\top\\\bm{d}_2^\top\\\vdots\\\bm{d}_N^\top\end{pmatrix}\in\mathbb{R}^{N\times N},\quad  D_0\equiv(a_{i,j})_{N\times N}=:\begin{pmatrix} \bm{a}_1^\top\\\bm{a}_2^\top\\\vdots\\\bm{a}_N^\top\end{pmatrix}\in\mathbb{R}^{N\times N}.
\end{equation}
Recall from \eqref{D1} that $D_1\equiv(b_{i,j})_{N\times N}=\bm{\beta}\bm{\kappa}^\top\in\mathbb{R}^{N\times N}$. 
According to the physical meaning of $D_0$, $D_1$ and $D$, we have, for $1\leq i,j\leq N$,
\begin{equation}
    a_{i,j}\geq0\;(i\neq j);\;a_{i,i}\leq0;\;\beta_{i}\geq0; \kappa_i\geq 0; \;d_{i,j}\geq0\;(i\neq j);\;d_{i,i}\leq0.
\end{equation}
Irreducibility of $D$ means there does not exist permutation matrix $P\in\mathbb{R}^{N\times N}$ such that 
\begin{equation}
    P^\top D P = \begin{pmatrix} C_1 & C_2 \\ \bm{0}_{N-r,r} & C_3 \end{pmatrix},  \quad 1 \leq r \leq N - 1,
\end{equation}
and $C_1$, $C_3$ are square matrices. 

Define $N_0:=\#\{i:1\leq i\leq N,\beta_i\neq0\}$.
Without loss of generality, assume $\beta_i>0$ for $1\leq i\leq N_0$ and $\beta_j=0$ for $N_0+1\leq j\leq N$ (when necessary, simultaneously permute rows and columns). 

Since $\text{rank}\,D_1=1$, $N_0\geq1$. Moreover, for $1\le i\le N_0$,
\begin{equation}
    \bm{a}_i^\top\bm{1}=\bm{d}_i^\top\bm{1}-\beta_i\bm{\kappa}^\top\bm{1}=-\beta_i\left(\bm{\kappa}^\top\bm{1}\right)=-\beta_i<0.
\end{equation}
If $N_0=N$, then $D_0$ is strictly row diagonally dominant and hence nonsingular by L\'{e}vy-Desplanques theorem. 

Now assume $N_0\leq N-1$. Denote
\begin{equation}
    D_0\equiv D-D_1=:\begin{pmatrix} A_1 & A_2 \\ A_3 & A_4 \end{pmatrix},
\end{equation}
where $A_1\in\mathbb{R}^{N_0\times N_0}$ and $A_4\in\mathbb{R}^{(N-N_0)\times(N-N_0)}$. 
Then 
\begin{equation}
    \begin{pmatrix} A_1&A_2\end{pmatrix}=\begin{pmatrix} \bm{a}_1^\top\\\bm{a}_2^\top\\\vdots\\\bm{a}_{N_0}^\top\end{pmatrix}=\begin{pmatrix} \bm{d}_1^\top\\\bm{d}_2^\top\\\vdots\\\bm{d}_{N_0}^\top\end{pmatrix}-[\beta_1,\beta_2,\dots,\beta_{N_0}]^\top\bm{\kappa}^\top=\begin{pmatrix} \bm{d}_1^\top-\beta_1\bm{\kappa}^\top\\\bm{d}_2^\top-\beta_2\bm{\kappa}^\top\\\vdots\\\bm{d}_{N_0}^\top-\beta_{N_0}\bm{\kappa}^\top\end{pmatrix}.
\end{equation}
Since $D$ is an irreducible $Q$-matrix, $\bm{d}_i^\top\bm{1}=0$ for $1\leq i\leq N$ and diagonal entries $d_{i,i}\;(1\leq i\leq N)$ are strictly negative. 
Because $D_0$ has nonnegative off-diagonal entries, it follows that $\bm{a}_i^\top\bm{1}<0$ for $1\leq i\leq N_0$, so $A_1$ is strictly diagonally dominant by rows and hence nonsingular by L\'{e}vy-Desplanques theorem.

Irreducibility of $D$ implies $A_3\neq\bm{0}_{(N-N_0)\times N_0}$. 
Define 
\begin{equation}
    N_1:=\#\{i:1\leq i\leq N-N_0,\;(A_3)_{i,:}\neq\bm{0}_{1\times N_0}\}.
\end{equation}
Note that $N_1\geq 1$. Without loss of generality, assume the first $N_1$ rows of $A_3$ are nonzero vectors (apply permutation to row and column simultaneously when necessary). Fix $1\leq j_0\leq N_1$. For the row $\bm{a}_{N_0+j_0}^\top$, choose $i_0\in\{1,\cdots,N_0\}$ with $a_{N_0+j_0,i_0}>0$ (such $i_0$ exists since the corresponding row of $A_3$ is nonzero). 

Perform the elementary row operation to the $(N_0+j_0)$-th row and obtain 
\begin{equation}\label{rowoperation1}
    \widehat{\bm{a}}_{N_0+j_0}^\top:=\bm{a}_{N_0+j_0}^\top-\left(\frac{a_{N_0+j_0,i_0}}{a_{i_0,i_0}}\right)\bm{a}_{i_0}^\top.
\end{equation}
Moreover,
\begin{equation}
    \widehat{\bm{a}}_{N_0+j_0}^\top\bm{1}=\left[\bm{a}_{N_0+j_0}^\top-\left(\frac{a_{N_0+j_0,i_0}}{a_{i_0,i_0}}\right)\bm{a}_{i_0}^\top\right]\bm{1}=-\left(\frac{a_{N_0+j_0,i_0}}{a_{i_0,i_0}}\right)\bm{a}_{i_0}^\top\bm{1}<0,
\end{equation}
and the off-diagonal entries of $\widehat{\bm{a}}_{N_0+j_0}^\top$ remain nonnegative; hence its diagonal entry is strictly negative and the row is strictly diagonally dominant. Applying this to all $1\leq j_0\leq N_1$ yields the matrix
\begin{equation}\label{matrix}
\begin{pmatrix}
\bm{a}_1&\bm{a}_2&\cdots&\bm{a}_{N_0}&\widehat{\bm{a}}_{N_0+1}&\cdots&\widehat{\bm{a}}_{N_0+N_1}&\bm{a}_{N_0+N_1+1}&\cdots&\bm{a}_{N}
\end{pmatrix}^\top=:
\begin{pmatrix} B_1 & B_2 \\ B_3 & B_4 \end{pmatrix},
\end{equation}
where $B_1\in\mathbb{R}^{(N_0+N_1)\times(N_0+N_1)}$ is strictly row diagonally dominant and thus nonsingular by L\'{e}vy-Desplanques theorem.
Note that elementary row (or column) operations preserve invertibility.
The original $(N_0+N_1)$-th leading principal submatrix $D_0\;(1:N_0+N_1,1:N_0+N_1)$ is also nonsingular. 

If $N-N_0-N_1=0$, $D_0$ is nonsingular. Now suppose $N-N_0-N_1\geq 1$. 

Denote
\begin{equation}
\begin{pmatrix}B_3&B_4\end{pmatrix}=\begin{pmatrix}
    \bm{a}_{N_0+N_1+1}^\top\\\vdots\\\bm{a}_{N}^\top
\end{pmatrix}=\begin{pmatrix}
    \bm{d}_{N_0+N_1+1}^\top\\\vdots\\\bm{d}_{N}^\top
\end{pmatrix}=:
\begin{pmatrix} \bm{0}_{(N-N_0-N_1)\times N_0} & B_5&B_4 \end{pmatrix},
\end{equation}
where irreducibility ensures $B_5\neq\bm{0}_{(N-N_0-N_1)\times N_1}$. 

Define 
\begin{equation}
    N_2:=\#\{i:1\leq i\leq N-N_0-N_1,\;(B_5)_{i,:}\neq\bm{0}_{1\times N_1}\}.
\end{equation}
Note that $N_2\geq 1$. Without loss of generality, assume only the first $N_2$ rows of $B_5$ are nonzero vectors (apply permutation to row and column simultaneously when necessary).
The same argument shows that the strictly larger leading principal submatrix $D_0(1:N_0+N_1+N_2,1:N_0+N_1+N_2)$ is nonsingular.

By induction, $D_0$ is nonsingular.

\section{Derivation of Fano Factor \eqref{fano} and Noise \eqref{cv}}\label{app8}
By definition and our results in \eqref{binomial}
\begin{equation}
\begin{aligned}
    FF:&=\frac{\sum_{n=0}^\infty n^2\mathcal{P}_n-(\sum_{n=0}^\infty n\mathcal{P}_n)^2}{\sum_{n=0}^\infty n\mathcal{P}_n}\\&=\frac{\sum_{n=0}^\infty n(n-1)\mathcal{P}_n+\sum_{n=0}^\infty n\mathcal{P}_n-(\sum_{n=0}^\infty n\mathcal{P}_n)^2}{\sum_{n=0}^\infty n\mathcal{P}_n}\\&
    =\frac{2B_2+B_1-B_1^2}{B_1}\\&
    =\frac{(1/\delta)\bm{\pi}^\top D_1(\delta\bm{I}_N-D)^{-1}D_1\bm{1}+(1/\delta)\bm{\pi}^\top D_1\bm{1}-(1/\delta^2)(\bm{\pi}^\top D_1\bm{1})^2}{(1/\delta)\bm{\pi}^\top D_1\bm{1}}\\&
    =\frac{\bm{\pi}^\top D_1(\delta\bm{I}_N-D)^{-1}D_1\bm{1}}{\bm{\pi}^\top D_1\bm{1}}+1-\frac{\bm{\pi}^\top D_1\bm{1}}{\delta}.
\end{aligned}
\end{equation}
Similarly, 
\begin{equation}
\begin{aligned}
\eta^2:&=\frac{\sum_{n=0}^\infty n^2\mathcal{P}_n-(\sum_{n=0}^\infty n\mathcal{P}_n)^2}{(\sum_{n=0}^\infty n\mathcal{P}_n)^2}\\&
=\frac{\sum_{n=0}^\infty n(n-1)\mathcal{P}_n+\sum_{n=0}^\infty n\mathcal{P}_n-(\sum_{n=0}^\infty n\mathcal{P}_n)^2}{(\sum_{n=0}^\infty n\mathcal{P}_n)^2}\\&
=\frac{2B_2+B_1-B_1^2}{B_1^2}\\&
=\frac{(1/\delta)\bm{\pi}^\top D_1(\delta\bm{I}_N-D)^{-1}D_1\bm{1}+(1/\delta)\bm{\pi}^\top D_1\bm{1}-(1/\delta^2)(\bm{\pi}^\top D_1\bm{1})^2}{(1/\delta^2)(\bm{\pi}^\top D_1\bm{1})^2}\\&
=\frac{\delta\bm{\pi}^\top D_1(\delta\bm{I}_N-D)^{-1}D_1\bm{1}}{(\bm{\pi}^\top D_1\bm{1})^2}+\frac{\delta}{\bm{\pi}^\top D_1\bm{1}}-1.
\end{aligned}
\end{equation}
\section{Proof of Inequality \eqref{neq}}\label{app9}
For discrete random variable $\{\mathcal{P}_n\}_{n\in\mathbb{N}}$, the moment generating function
\begin{equation}
    \mathcal{M}[\{\mathcal{P}_n\}_{n\in\mathbb{N}}](t):=\sum_{n=0}^\infty \mathcal{P}_n\mathrm{e}^{nt}.
\end{equation}
Using \eqref{Bound1}, for $t>0$, 
\begin{equation}
\begin{aligned}
    \mathcal{M}\left[\{\mathcal{P}_n\}_{n\in\mathbb{N}}\right](t)&=\sum_{n=0}^\infty \mathcal{P}_n\mathrm{e}^{nt}\\
    &\leq \sum_{n=0}^\infty\mathrm{e}^{\lVert D_1\rVert_{\infty}/\delta}\frac{1}{n!}\left(\frac{\lVert D_1\rVert_{\infty}}{\delta}\right)^n\mathrm{e}^{nt}\\
    &=\mathrm{e}^{2\lVert D_1\rVert_{\infty}/\delta}\sum_{n=0}^\infty\mathrm{e}^{-\lVert D_1\rVert_{\infty}/\delta}\frac{1}{n!}\left(\frac{\lVert D_1\rVert_{\infty}}{\delta}\right)^n\mathrm{e}^{nt}\\
    &=\mathrm{e}^{2\lVert D_1\rVert_{\infty}/\delta}\times\mathcal{M}\left[\text{Pois}(\frac{\lVert D_1\rVert_\infty}{\delta})\right](t)\\&<\infty. 
\end{aligned}
\end{equation}
\newpage

\bibliographystyle{siam}
\bibliography{BIB}

\begin{thebibliography}{10}

\bibitem{BinomialMomentEquations2011Phys.Rev.Lett.}
{\sc B.~Barzel}, {\em Binomial moment equations for stochastic reaction systems}, Phys. Rev. Lett., 106 (2011).

\bibitem{Stochasticanalysiscomplex2012Phys.Rev.Ea}
{\sc B.~Barzel and O.~Biham}, {\em Stochastic analysis of complex reaction networks using binomial moment equations}, Phys. Rev. E, 86 (2012), p.~031126.

\bibitem{StochasticProcessesCell2021}
{\sc P.~C. Bressloff}, {\em Stochastic {{Processes}} in {{Cell Biology}}: {{Volume I}}}, vol.~41 of Interdisciplinary {{Applied Mathematics}}, Springer, Cham, 2021.

\bibitem{Exactdistributionsstochastic2022Phys.Rev.E}
{\sc M.~Chen, S.~Luo, M.~Cao, C.~Guo, T.~Zhou, and J.~Zhang}, {\em Exact distributions for stochastic gene expression models with arbitrary promoter architecture and translational bursting}, Phys. Rev. E, 105 (2022), p.~014405.

\bibitem{MechanismTranscriptionalBursting2014Cell}
{\sc S.~Chong, C.~Chen, H.~Ge, and X.~S. Xie}, {\em Mechanism of transcriptional bursting in bacteria}, Cell, 158 (2014), pp.~314--326.

\bibitem{CourseProbabilityTheory2000}
{\sc K.~L. Chung}, {\em A {{Course}} in {{Probability Theory}}}, Academic Press, San Diego, 2000.

\bibitem{AppliedStochasticAnalysis2019}
{\sc W.~E, T.~Li, and E.~{Vanden-Eijnden}}, {\em Applied {{Stochastic Analysis}}}, American Mathematical Society, Providence, 2019.

\bibitem{generalmethodnumerically1976J.Comput.Phys.}
{\sc D.~T. Gillespie}, {\em A general method for numerically simulating the stochastic time evolution of coupled chemical reactions}, J. Comput. Phys., 22 (1976), pp.~403--434.

\bibitem{StochasticSimulationChemical2007Annu.Rev.Phys.Chem.}
{\sc D.~T. Gillespie}, {\em Stochastic simulation of chemical kinetics}, Annu. Rev. Phys. Chem., 58 (2007), pp.~35--55.

\bibitem{MatrixComputations2013}
{\sc G.~H. Golub and C.~F.~V. Loan}, {\em Matrix {{Computations}}}, Johns Hopkins University Press, Baltimore, 2013.

\bibitem{ExtrinsicNoiseHeavyTailed2020Phys.Rev.Lett.}
{\sc L.~Ham, R.~D. Brackston, and M.~P.~H. Stumpf}, {\em Extrinsic noise and {{Heavy-Tailed Laws}} in gene expression}, Phys. Rev. Lett., 124 (2020), p.~108101.

\bibitem{stochasticvsdeterministic2024Nat.Commun.}
{\sc L.~Ham, M.~A. Coomer, K.~{\"O}cal, R.~Grima, and M.~P.~H. Stumpf}, {\em A stochastic vs deterministic perspective on the timing of cellular events}, Nat. Commun., 15 (2024), p.~5286.

\bibitem{Exactlysolvablemodels2020J.Chem.Phys.}
{\sc L.~Ham, D.~Schnoerr, R.~D. Brackston, and M.~P.~H. Stumpf}, {\em Exactly solvable models of stochastic gene expression}, J. Chem. Phys., 152 (2020), p.~144106.

\bibitem{StochasticGeneExpression2019SIAMJ.Appl.Math.}
{\sc U.~Herbach}, {\em Stochastic gene expression with a multistate promoter: breaking down exact distributions}, SIAM J. Appl. Math., 79 (2019), pp.~1007--1029.

\bibitem{Stochasticmodelingautoregulatory2020Biophys.J.a}
{\sc J.~Holehouse, Z.~Cao, and R.~Grima}, {\em Stochastic modeling of autoregulatory genetic feedback loops: a review and comparative study}, Biophys. J., 118 (2020), pp.~1517--1525.

\bibitem{MatrixAnalysis2012}
{\sc R.~A. Horn and C.~R. Johnson}, {\em Matrix {{Analysis}}}, Cambridge University Press, New York, 2012.

\bibitem{MultimodalityFlexibilityStochastic2013Bull.Math.Biol.}
{\sc G.~d. C.~P. Innocentini, M.~Forger, A.~F. Ramos, O.~Radulescu, and J.~E.~M. Hornos}, {\em Multimodality and flexibility of stochastic gene expression}, Bull. Math. Biol., 75 (2013), pp.~2600--2630.

\bibitem{Smallproteinnumber2020J.Chem.Phys.}
{\sc C.~Jia and R.~Grima}, {\em Small protein number effects in stochastic models of autoregulated bursty gene expression}, J. Chem. Phys., 152 (2020), p.~084115.

\bibitem{AnalyticalTimeDependentDistributions2023SIAMJ.Appl.Math.}
{\sc C.~Jia and Y.~Li}, {\em Analytical time-dependent distributions for gene expression models with complex promoter switching mechanisms}, SIAM J. Appl. Math., 83 (2023), pp.~1572--1602.

\bibitem{IntrinsicNoiseStochastic2011Phys.Rev.Lett.}
{\sc T.~Jia and R.~V. Kulkarni}, {\em Intrinsic noise in stochastic models of gene expression with molecular memory and bursting}, Phys. Rev. Lett., 106 (2011), p.~058102.

\bibitem{Whatcanwe2024PLoSComput.Biol.}
{\sc F.~Jiao, J.~Li, T.~Liu, Y.~Zhu, W.~Che, L.~Bleris, and C.~Jia}, {\em What can we learn when fitting a simple telegraph model to a complex gene expression model?}, PLoS Comput. Biol., 20 (2024), p.~e1012118.

\bibitem{StochasticityTranscriptionalRegulation2001Biophys.J.}
{\sc T.~B. Kepler and T.~C. Elston}, {\em Stochasticity in transcriptional regulation: origins, consequences, and mathematical representations}, Biophys. J., 81 (2001), pp.~3116--3136.

\bibitem{Inferringkineticsstochastic2013GenomeBiol.}
{\sc J.~K. Kim and J.~C. Marioni}, {\em Inferring the kinetics of stochastic gene expression from single-cell {{RNA-sequencing}} data}, Genome Biol., 14 (2013), p.~R7.

\bibitem{Centraldogmasinglemolecule2011Nature}
{\sc G.-W. Li and X.~S. Xie}, {\em Central dogma at the single-molecule level in living cells}, Nature, 475 (2011), pp.~308--315.

\bibitem{$GI^Xinfty$system1990J.Appl.Probab.}
{\sc L.~Liu, B.~R.~K. Kashyap, and J.~G.~C. Templeton}, {\em On the ${{GI}}^{{X}}/{{G}}/\infty$ system}, J. Appl. Probab., 27 (1990), pp.~671--683.

\bibitem{Inferringtranscriptionalbursting2023R.Soc.OpenSci.}
{\sc S.~Luo, Z.~Zhang, Z.~Wang, X.~Yang, X.~Chen, T.~Zhou, and J.~Zhang}, {\em Inferring transcriptional bursting kinetics from single-cell snapshot data using a generalized telegraph model}, R. Soc. Open Sci., 10 (2023), p.~221057.

\bibitem{AnalysisInfiniteServerQueue2002QueueingSyst.}
{\sc H.~Masuyama and T.~Takine}, {\em Analysis of an infinite-server queue with batch {{Markovian}} arrival streams}, Queueing Syst., 42 (2002), pp.~269--296.

\bibitem{GillesPy2BiochemicalModeling2023Lett.Biomath.}
{\sc S.~Matthew, F.~Carter, J.~Cooper, M.~Dippel, E.~Green, S.~Hodges, M.~Kidwell, D.~Nickerson, B.~Rumsey, J.~Reeve, L.~R. Petzold, K.~R. Sanft, and B.~Drawert}, {\em {{GillesPy2}}: a biochemical modeling framework for simulation driven biological discovery}, Lett. Biomath., 10 (2023), pp.~87--103.

\bibitem{finitestateprojection2006J.Chem.Phys.}
{\sc B.~Munsky and M.~Khammash}, {\em The finite state projection algorithm for the solution of the chemical master equation}, J. Chem. Phys., 124 (2006), p.~044104.

\bibitem{UsingGeneExpression2012Science}
{\sc B.~Munsky, G.~Neuert, and A.~{van Oudenaarden}}, {\em Using gene expression noise to understand gene regulation}, Science, 336 (2012), pp.~183--187.

\bibitem{$Minfty$queue1986J.Appl.Probab.}
{\sc C.~A. O'Cinneide and P.~Purdue}, {\em The ${{M}}/{{M}}/\infty$ queue in a random environment}, J. Appl. Probab., 23 (1986), pp.~175--184.

\bibitem{Regulationnoiseexpression2002NatureGenet.}
{\sc E.~M. Ozbudak, M.~Thattai, I.~Kurtser, A.~D. Grossman, and A.~{van Oudenaarden}}, {\em Regulation of noise in the expression of a single gene}, Nature Genet., 31 (2002), pp.~69--73.

\bibitem{Summingnoisegene2004Nature}
{\sc J.~Paulsson}, {\em Summing up the noise in gene networks}, Nature, 427 (2004), pp.~415--418.

\bibitem{Modelsstochasticgene2005Phys.LifeRev.}
\leavevmode\vrule height 2pt depth -1.6pt width 23pt, {\em Models of stochastic gene expression}, Phys. Life Rev., 2 (2005), pp.~157--175.

\bibitem{MarkovianModelingGeneProduct1995Theor.Popul.Biol.}
{\sc J.~Peccoud and B.~Ycart}, {\em Markovian modeling of gene-product synthesis}, Theor. Popul. Biol., 48 (1995), pp.~222--234.

\bibitem{DistanceMattersImpact2013Phys.Rev.Lett.}
{\sc O.~Pulkkinen and R.~Metzler}, {\em Distance matters: {{The}} impact of gene proximity in bacterial gene regulation}, Phys. Rev. Lett., 110 (2013), p.~198101.

\bibitem{NoiseGeneExpression2005Science}
{\sc J.~M. Raser and E.~K. O'Shea}, {\em Noise in gene expression: origins, consequences, and control}, Science, 309 (2005), pp.~2010--2013.

\bibitem{Approximationinferencemethods2017J.Phys.A-Math.Theor.}
{\sc D.~Schnoerr, G.~Sanguinetti, and R.~Grima}, {\em Approximation and inference methods for stochastic biochemical kinetics---a tutorial review}, J. Phys. A-Math. Theor., 50 (2017), p.~093001.

\bibitem{Analyticaldistributionsstochastic2008Proc.Natl.Acad.Sci.U.S.A.}
{\sc V.~Shahrezaei and P.~S. Swain}, {\em Analytical distributions for stochastic gene expression}, Proc. Natl. Acad. Sci. U. S. A., 105 (2008), pp.~17256--17261.

\bibitem{QueuingModelsGene2020Biophys.J.}
{\sc C.~Shi, Y.~Jiang, and T.~Zhou}, {\em Queuing models of gene expression: analytical distributions and beyond}, Biophys. J., 119 (2020), pp.~1606--1616.

\bibitem{NascentRNAkinetics2024Phys.Rev.E}
{\sc C.~Shi, X.~Yang, T.~Zhou, and J.~Zhang}, {\em Nascent {{RNA}} kinetics with complex promoter architecture: analytic results and parameter inference}, Phys. Rev. E, 110 (2024), p.~034413.

\bibitem{MammalianGenesAre2011Science}
{\sc D.~M. Suter, N.~Molina, D.~Gatfield, K.~Schneider, U.~Schibler, and F.~Naef}, {\em Mammalian genes are transcribed with widely different bursting kinetics}, Science, 332 (2011), pp.~472--474.

\bibitem{Solvingstochasticgeneexpression2024Biophys.J.}
{\sc J.~{Szavits-Nossan} and R.~Grima}, {\em Solving stochastic gene-expression models using queueing theory: a tutorial review}, Biophys. J., 123 (2024), pp.~1034--1057.

\bibitem{StochasticProcessesPhysics2007}
{\sc N.~G. van Kampen}, {\em Stochastic {{Processes}} in {{Physics}} and {{Chemistry}}}, Elsevier, Amsterdam, 2007.

\bibitem{Poissonrepresentationbridge2023J.R.Soc.Interface}
{\sc X.~Wang, Y.~Li, and C.~Jia}, {\em Poisson representation: a bridge between discrete and continuous models of stochastic gene regulatory networks}, J. R. Soc. Interface, 20 (2023), p.~20230467.

\bibitem{momentconvergencemethodstochastic2016J.Chem.Phys.a}
{\sc J.~Zhang, Q.~Nie, and T.~Zhou}, {\em A moment-convergence method for stochastic analysis of biochemical reaction networks}, J. Chem. Phys., 144 (2016).

\bibitem{PromotermediatedTranscriptionalDynamics2014Biophys.J.}
{\sc J.~Zhang and T.~Zhou}, {\em Promoter-mediated transcriptional dynamics}, Biophys. J., 106 (2014), pp.~479--488.

\bibitem{Analyticalresultsmultistate2012SIAMJ.Appl.Math.}
{\sc T.~Zhou and J.~Zhang}, {\em Analytical results for a multistate gene model}, SIAM J. Appl. Math., 72 (2012), pp.~789--818.

\end{thebibliography}

\end{document}